\documentclass[11pt]{article}
\usepackage{graphicx}
\usepackage{tikz}
\usepackage[english]{babel}
\usepackage[T1]{fontenc}
\usepackage{hyperref}
\usepackage{amstext,amsmath}
\usepackage{dcolumn,booktabs}
\usepackage{amsfonts}
\usepackage{float}
\usepackage{comment}
\usepackage{listings}
\usepackage{url}
\usepackage{longtable}
\usepackage[tableposition=b]{caption}
\usepackage{color}
\usepackage{graphicx}
\usepackage[margin=1in]{geometry}
\usepackage{subcaption}
\usepackage{amssymb}
\usepackage{etex}
\usepackage{pstricks}
\usepackage{pst-plot}
\usepackage{pstricks-add}
\usepackage[round]{natbib}   
\usepackage[T1]{fontenc}
\usepackage{setspace}
\usepackage{bbm}
\usepackage{graphicx,floatpag,fancyhdr}
\fancypagestyle{plainlower}{	
	\fancyhf{}
	\fancyfoot[C]{\raisebox{-6\baselineskip}{\thepage}}
	
}
\usepackage{lscape} 
\usepackage{placeins}
\usepackage{tabularx}

\newlength{\bibitemsep}
\newlength{\bibparskip}
\let\oldthebibliography\thebibliography
\renewcommand\thebibliography[1]{%
	\oldthebibliography{#1}%
	\setlength{\parskip}{\bibitemsep}%
	\setlength{\itemsep}{\bibparskip}%
}

\usepackage{listings}
\usepackage[toc,page]{appendix}
\defcitealias{HBR2014}{HBR, 2014}
\defcitealias{hkp2023}{hkp, 2023}
\defcitealias{hkp}{hkp, 2021}
\usepackage{amsfonts, amsmath, amsthm, amssymb}

\theoremstyle{definition}

\usepackage{hyperref}
\begin{document}
		\title{The Broken Rung: Gender and the Leadership Gap}
\author{Ingrid Haegele\footnote{Ludwig-Maximilians University; ingrid.haegele@econ.lmu.de.  This paper has benefited tremendously from comments and suggestions from  Sydnee Caldwell, David Card, Supreet Kaur, Patrick Kline, Conrad Miller, Derek Neal, Evan Rose, Ricardo Perez-Truglia, Fabian Waldinger,  Christopher Walters,  Andrea Weber,  and Francis Wong. I thank the employees of my institutional partner for their continuing support. This work was supported by the Alfred P. Sloan Foundation Pre-doctoral Fellowship on the Economics of an Aging Workforce awarded through the NBER,  IRLE Student Research Grant, O-LAB Labor Science Initiative Grant, Strandberg Grant for Gender in Economic Research, UC Berkeley Matrix Dissertation Fellowship, and Washington Center for Equitable Growth Doctoral Grant.}}
\date{April 2024}
\maketitle

\bigskip

\begin{abstract}
\noindent Addressing female underrepresentation in leadership positions has become a key policy objective. However, little is known about the extent to which leadership appeals differently to women. Collecting new data from a large firm, I document that women are substantially less likely to apply for early-career promotions.  Realized application patterns and large-scale surveys reveal the role of an understudied feature of promotions---having to assume responsibility over a team---which is less appealing to women. This gender difference is not accounted for by standard explanations, such as success likelihood or confidence, but is rather a product of common design features of leadership positions. 


\end{abstract}
\clearpage
\newpage

\section{Introduction}
Much attention has been devoted to the fact that women are less likely to hold leadership positions than men. In past decades, policymakers, organizations, and researchers have invested substantial resources to address demand-side factors that may hinder women's career progression, for instance by implementing female quotas for top positions or by debiasing hiring practices in organizations (\citealp{OECD2020}). However, a growing body of evidence points to meaningful gender differences in labor supply decisions that may also contribute to the representation gap  (\citealp{Hospido2019}, \citealp{Fluchtmann2021}).  While previous work in economics has documented that characteristics, such as  flexibility or commutes have differential effects on women's labor supply decisions (\citealp{MasPallais2017}, \citealp{WiswallZafar2018}, \citealp{LeBarbanchon2021}), little is known about the extent to which the process of climbing the leadership ladder---and the changes that this may bring with it---is less appealing to women relative to men.

Because of data limitations, promotions to higher-level positions are often measured based on pay increases or flows between occupations, making it difficult to infer why climbing the leadership ladder may be unappealing. Laboratory studies have found evidence for gender differences in preferences for different dimensions of leadership, such as the desire for power or authority in decision-making (\citealp{ertac}).
However, relatively little emphasis has been placed on one particularly salient dimension of leadership positions in real workplaces: responsibility over a team.\footnote{While some laboratory studies have focused on the effect of team composition on gender differences in willingness to lead, leadership in such studies can only be approximated by actions such as making a decision for the group (\citealp{eckel}).} This lack of emphasis is in stark contrast with the fact that in many firms, taking on responsibility over a team is a key prerequisite for career advancement (\citetalias{hkp2023}), and with anecdotal and survey evidence suggesting that having to lead others is  often negatively perceived.\footnote{In a survey of US workers, 66\% of workers report not wanting to lead others (\citetalias{HBR2014}). Reports from HR practitioners and in online worker forums indicate that many workers perceive team responsibility as being burdensome (\citealp{Inhersight}).}  

This paper provides the first evidence that having to assume responsibility over a team is a key reason why women are less likely to seek early-career promotions that allow them to climb the leadership ladder. I collect a new dataset that combines the universe of job application and vacancy data of a large firm with detailed personnel records and survey responses, spanning over 30,000 white-collar and management employees from 2015 to 2019. Leveraging detailed measures of job hierarchy,  I document the existence of a broken rung:  women  in lower-level positions are substantially less likely to apply for early-career promotions than men. Both realized application patterns and large-scale surveys at the firm point to the role of team leadership, which is a common yet understudied feature of promotions, as key driver of the gender application gap.  The documented gender differences with respect to team leadership are not fully explained by standard factors, such as family constraints, confidence, and differences in perceived success likelihood.  Instead, I document that the job features that team leadership positions typically entail have differential impacts on women relative to men. 
 
The data in this study come from a large multinational firm that is one of the largest manufacturers in Europe, employing over 200,000 workers. To examine internal career progression to higher-level positions, I focus my analysis on the firm's largest internal labor market, consisting of over 30,000 white-collar and management employees in Germany. The firm's workforce is comparable to that of other large German firms in terms of demographics and female representation and spans a broad set of positions, with female shares varying between 9\% in engineering and 70\% in HR. A key advantage of my setting is that the firm---like many other large organizations---requires employees to actively apply in order to make internal job switches, including promotions (\citetalias{hkp}). This feature allows me to analyze employees' application decisions independent of their hiring outcomes. These application decisions are made under uncertainty, with only 27\% of internal applications being successful and the majority of applications made to positions outside of employees' current team. 

The objective of this study is to test whether men and women differ in the extent to which they seek to climb the leadership ladder.  A canonical view in economics is that such increases in job hierarchy involve moving to positions with increased authority (\citealp{Rosen1982}). In practice, such increases in authority can take different forms, for instance by taking on responsibility over people, projects, or business decisions  (\citealp{SHRM2021}).  Capturing granular differences in authority at the individual level is empirically challenging because most common datasets do not contain information on job responsibilities.  This is particularly difficult at lower parts of the leadership ladder, where position titles often only make functional distinctions (e.g., marketing vs engineering) and do not distinguish a job's degree of responsibility (e.g., individual contributor vs leader of a small team). Distinguishing differences at early leadership levels,  however, is of particular interest when studying the leadership gap. Because firms prioritize internal candidates when filling higher-level positions, whether or not employees enter the leadership pipeline has long-term career impacts (\citetalias{hkp2023}). In addition, a growing body of evidence documents that important gender differences with respect to career progression emerge early in employees' careers, further motivating a focus on lower parts of the leadership ladder (\citealp{BronsonThoursie2019}, \citealp{Hospido2019}, \citealp{Friebel2022}, \citealp{azmat2020gender}). 

In order to identify promotions as transitions to higher-level positions for employees at any part of the leadership ladder,  I collect detailed information on relevant dimensions of job authority. First, I use the number of direct reports to capture the extent of responsibility an individual has over others (e.g., with respect to hiring, firing, and performance review). Second, I use the firm's measure of managerial autonomy, which captures the extent to which an employee has autonomy over working hours and business decisions.  Third, I collect information on reporting relationships in order to capture an employee's authority as measured by their reporting distance to the CEO. Because these measures of job responsibility may differ across areas in the firm, and because men and women differentially sort across these areas, my preferred approach of identifying higher-level positions combines all three measures into a one-dimensional ranking.\footnote{For instance, positions may differ in the number of reports not only because of the extent of responsibility they entail, but also because team size varies across areas of the firm (e.g., teams in engineering are generally larger than in finance). Also accounting for the extent of managerial autonomy and the reporting distance to the CEO therefore enables more accurate comparisons of job authority across different areas.} Specifically, I use the first principal component of these three dimensions of job authority, which explains 61\% of the variation and provides a consistent ordering of all positions at the firm. To test the robustness of my findings, I use pay increases as an alternative measure for identifying promotions, which captures differences in job authority to the extent to which they are remunerated differently.

My findings reveal the existence of a broken rung. Women in lower-level positions are substantially less likely to apply for early-career promotions relative to their male counterparts. When using my preferred approach to identify higher-level positions based on combining the  three dimensions of job authority, I find a gender application gap of 27.4\%.  However, women who already hold a leadership position are not less likely to apply for subsequent promotions, indicating that early-career promotions are of particular importance for understanding gender differences with respect to leadership. My results are robust to alternative measures of promotion. When using pay increases to identify promotions, I also find that women in lower-level positions, but not those at high rungs of the leadership ladder, are less likely to apply for promotions.

Why are women in lower-level positions less likely to apply for promotions? One potential explanation of the observed application gap is that it is driven by differences in worker characteristics---such as hours constraints, preferences for location, or confidence---rather than specific features of promotions. While prior work has documented the general importance of such worker-level factors for gender differences in labor supply decisions (\citealp{MasPallais2017}, \citealp{WiswallZafar2018}, \citealp{Coffman2020}, \citealp{Fluchtmann2021}), I do not find that these factors can account for the differences in applications for promotions I document. For instance, I find large application differences even among employees for whom hours constraints are less binding or who are particularly confident. Similarly, large gender differences in applications exist even for promotions to positions that are located in employees' current city. Moreover, the observed application gap is not explained by differences in perceived success likelihood or differential access to information about job openings. 

My findings indicate that the requirement to assume responsibility over a team---which is a common feature of promotions---makes higher-level positions less appealing to women. Responses to a large-scale survey of the firm's employees suggest that responsibility for a team is a particularly salient dimension of early-career promotions. The survey invited all employees in my sample to participate and received a 50\% response rate. Employees who already hold leadership positions were asked about the most salient changes in job characteristics that accompanied their first promotion. Having to take on responsibility over a team is reported almost twice as often as any other feature of promotions, such as greater responsibility over projects or business decisions. While a growing body of experimental research emphasizes different dimensions of leadership that may give rise to gender differences (\citealp{eckel}), the extent to which responsibility over a team is less appealing to women and its role for early-career promotions in the context of real-world settings has received relatively little attention. 
 
When asked where employees would like to see themselves with respect to their career progression, women in lower-level positions were 36\% less likely  to report wanting to take on responsibility over a team. This difference is large in magnitude and is not accounted for by standard explanations, such as family status, risk preferences, or confidence. I also find that this gender leadership gap is similar in magnitude across different survey waves at the firm: an earlier survey of employees in Germany found a gap of 32\%, while the corresponding gap among the firm's employees in the US was 39\%.  In addition, the documented gender leadership gap is robust across different survey questions. A supplementary question prompted respondents to indicate which of the workshops that the firm considers offering they would like to participate in, while informing them that workshops that received little interest may not be offered. Women were 23\% less likely to choose a workshop on how to become a successful team leader. Similarly, even when holding constant other job characteristics of team leadership positions in a hypothetical job choice question, women were 13\% less likely to choose a position with team responsibility over a position without such responsibility.  These findings provide the first evidence that meaningful gender differences arise with respect to taking on responsibility over a team.

Employees' realized application decisions confirm that the reported gender gap with respect to team leadership translates into differences in applications. Leveraging the fact that not all promotions require taking on responsibility over a team, I compare realized applications for team leader positions to applications to other types of higher-level positions, such as those requiring taking on responsibility over a project or product. While men are more than twice as likely to apply for a promotion if it requires taking on responsibility over a team, this is not true for women, yielding a large application gap for promotions that involve responsibility over a team. While such team leadership positions may be different from other promotions, I do not find that standard vacancy characteristics---including the extent of listed skill requirements, the gender composition of peers and supervisors, or how attractive the job posting appears---can explain my results. Together, the complementary findings from the employee survey and the realized applications provide strong evidence in support of meaningful gender differences with respect to team leadership. 

 Why is team leadership less appealing to women and what does this imply for organizations seeking to increase female representation in higher-level positions? Anecdotal and survey evidence across different organizations suggests that having to lead others is generally negatively perceived by many employees (\citetalias{HBR2014}, \citealp{Inhersight}). Survey responses from employees in my sample corroborate this finding and point to the role of two common position features of team leadership that are particularly unappealing: the administrative tasks related to having to lead large teams and the requirement to resolve conflict in the team. Using hypothetical job choice questions, I find that women are 39\% more willing to trade off higher pay to avoid leading a large team, which carries a higher administrative burden. In addition, women are 14\% more likely to forgo higher pay in order to avoid leading an unknown team with a larger potential for conflict. I also find that women in lower-level positions have more negative perceptions regarding team leadership. Female employees are 29\% more likely to overestimate the size of teams and 12\% more likely to overestimate the frequency of conflict that team leaders at the firm face. In contrast, I find no evidence that the actual experiences of female and male leaders differ with respect to these features. Taken together, these results suggest that women have overly negative perceptions of what team leadership entails and are more likely to be deterred by these features. In combination, these gender differences explain half of the application gap for team leadership positions, highlighting the importance of the design of team leadership positions for alleviating the gender leadership gap. 
 
 My results suggest that organizations may be able to increase female applications by better tailoring the design of leadership to women, for instance by offering a larger share of team leadership positions with smaller team sizes. In addition, offering more support and information for how to navigate the challenges related to leading a team may also improve the appeal of these positions. In the survey, women place a particularly high value on mentoring, and frequently express interest in learning how current leaders respond to conflict. Moreover, there appears to be scope for organizations to increase women's applications by assuaging overly negative perceptions of team leadership.  Strategies suggested in the survey include providing more information in job postings, such as information on the size and composition of the team, and access to job shadowing opportunities, which would offer an opportunity to directly observe the actual experience of team leaders.

This paper contributes to a large literature in economics that studies gender differences in labor market outcomes (\citealp{Goldin2014}, \citealp{BlauKahn2017}). A growing body of work has highlighted the importance of studying gender differences in career progression as a key contributor for the gender pay gap (\citealp{BronsonThoursie2019}, \citealp{CullenPerezTruglia}). Because promotions are difficult to measure outside of narrowly defined settings and are often accompanied by a variety of changes in job characteristics, little is known about whether and why climbing the leadership ladder may be less appealing to women. By combining insights from employees' realized applications with large-scale survey evidence, I provide the first evidence that the requirement to lead a team has differential effects on whether women and men seek early-career promotions. This finding documents an important reason for the importance of early-career gender gaps (\citealp{BronsonThoursie2019}, \citealp{Hospido2019}, \citealp{Friebel2022}, \citealp{azmat2020gender}, \citealp{Benson2021potential}).  Because women are less likely to enter the leadership pipeline and because organizations typically prioritize internal candidates when filling higher-level positions,  these early-career gaps can have cascading effects along the leadership ladder (\citetalias{hkp2023}).   

By documenting the important role of team responsibility as a salient but understudied dimension of promotions, this paper also contributes to the literature in economics that highlights the impact of job characteristics on workers' labor supply decisions (\citealp{MasPallais2017}, \citealp{WiswallZafar2018}, \citealp{wassermanhours},  \citealp{LeBarbanchon2021}, \citealp{FolkeRickne2022}). While previous work has mostly focused on job characteristics such as flexibility and commuting time, my findings imply that accounting for the extent to which positions require employees to lead a team is important for understanding and potentially alleviating gender differences in labor supply decisions.

This paper also provides a comprehensive analysis of the potential drivers of gender differences in application behavior across a wide range of occupations. Previous work has highlighted several reasons why women and men may differ in their decisions regarding job applications  (\citealp{Flory2014},  \citealp{Hospido2019},  \citealp{Fluchtmann2021}, \citealp{Cortes2022}, \citealp{AbrahamStein}, \citealp{Coffman2020}, \citealp{azmat2020gender}). By collecting an unusually rich dataset that combines the universe of application decisions  within a large organization with survey evidence on the potential reasons underlying employees’ application decisions, I am able to take a range of relevant mechanisms into account that may underlie gender differences in application behavior. My findings indicate that even for subgroups of women who are less likely to be affected by family constraints, who have high risk preferences, and those with high confidence, the requirement to take on responsibility over a team remains a major deterrent of applications for early-career promotions. This finding suggests that taking into account the design of leadership positions is important for identifying and alleviating gender differences in applications. 

The rest of the paper proceeds as follows. Sections \ref{sec:setting} and \ref{sec:data} introduce the setting and the new data I collect. Section \ref{sec:hierarchy} describes how I define and measure promotions along the leadership ladder. Section \ref{sec:applications} documents the existence of a large gender gap in applications for early-career promotions. Section \ref{sec:teamleadership} demonstrates that it is the requirement to take on responsibility over a team that makes promotions less appealing to women. Section \ref{sec:consequences} discusses implications for the design of team leadership positions. Section \ref{sec:conclusion} concludes.

\section{Setting}
\label{sec:setting}
This paper uses a unique combination of personnel records, job application data, and large-scale survey evidence from a large multinational firm. The firm employs over 200,000 workers around the world and is one of the largest manufacturers in Europe. To maintain confidentiality, I refrain from providing details that could be used to identify the firm. As a large manufacturer, the firm's internal labor market consists of over 200 different occupations. The majority of positions are in technical areas, such as engineering or production, which are traditionally male-dominated. However, the firm also employs more female-leaning occupations, such as HR, finance, and marketing, allowing me to analyze internal career outcomes across both male-leaning and female-leaning areas. Since the goal of this study is to analyze career progression to higher-level positions, I  restrict my analysis to white-collar and management employees at the firm (i.e. employees that are either already in or could ultimately attain management positions). While the firm operates in many different countries, Germany represents the largest internal labor market of the firm. I therefore focus my analysis on all 30,000 white-collar and management employees who are based in Germany. 

Table \ref{table:sample_descriptive} provides summary statistics for my analysis sample which consists of over 400,000 employee-by-quarter observations from 2015 to 2019.\footnote{This sample is similar to the one studied in \cite{Haegele2023}. To maintain confidentiality, I do not disclose the exact number of employees.} Women represent 20\% of employees in the sample, which is consistent with the underrepresentation of women in technical occupations. Because I restrict to white-collar and management employees with regular employment contracts (as opposed to those with marginal employment such as mini jobs), employee qualification in my sample is high. The average employee holds a Bachelor's degree and 93\% of employees work full-time.  Employee tenures at the firm tend to be long, with an average tenure of 13 years. These demographic patterns are comparable to other large manufacturing firms in Germany. In Appendix Table \ref{table:bibb_2018}, I compare employees in my sample to those  in large manufacturing firms in the BiBB, a representative survey of the German workforce conducted in 2018. I find very similar patterns with respect to most employee characteristics (e.g., gender, age, German citizenship, and martial status).

Entry into the firm is most common in lower-level positions where employees work as individual contributors.\footnote{In my sample, 95\% of individuals enter the firm as individual contributors in a lower-level position.} As in most organizations, HR policies stipulate that the usual next career step for employees in lower-level positions is a promotion to a first leadership level, which  involves taking on more job responsibility, for instance in terms of leadership over a team. Alternatively, increased job responsibility could stem from taking on a position as head of project or in an expert role, which do not involve responsibility over a team (\citealp{SHRM2021}).  All three of these types of higher-level positions are characterized by enhanced job responsibility in terms of authority relative to lower-level positions, but only  team leader positions require direct responsibility over a team. While at this first level of the leadership ladder, 79\% of positions require responsibility over a team,  team responsibility becomes increasingly important as hierarchy levels increase.  The vast majority (86\%) of executive positions and virtually all top management positions (98\%) require responsibility over a team. 

Men and women at the firm substantially differ in their career outcomes. Women earn on average 34\% less and are 81\% less likely to be in the top decile of earnings at the firm. They are also 54\% less likely to hold top management positions. When defining managers as those who have responsibility over a team, I find that women are 62\% less likely to hold these manager positions. Even among those who lead teams, women on average have 16\% fewer subordinates. These differences are similar to broader patterns in Germany. For instance, in the BiBB, women are 54\% less likely to hold top leadership positions, 23\% less likely to hold mid-level leadership positions, and 40\% less likely to hold lower-level leadership positions relative to base rates in West Germany. Moreover, the extent of female underrepresentation in Germany accords with that in other Western countries, such as the United States or other members of the European Union.\footnote{For instance, in 2019, the gender pay gap for full-time employees was 11\% in the European Union, 14\% in Germany, and 18\% in the United States (\citealp{OECD2022}). Both in Germany and the United States, women are underrepresented on corporate boards and hold only 29\% and 24\% of seats, respectively (\citealp{Deloitte2021}). Robustness exercises using data on the firm's employees who are based in the United States corroborate that the patterns I document also apply to settings outside Germany.}

Like many other large organizations, the firm requires employees to actively apply using an online job portal in order to make internal job transitions, including promotions. This feature enables me to isolate employees' application decisions from the firm's callback and hiring decisions. Employees can access every job opening at the firm through a centralized online job portal and are required to apply through the portal,  which typically takes less than five minutes to complete (see Appendix Figure \ref{fig:jobportal} for an illustration of the portal). Such active application systems are very common (\citetalias{hkp}). The firm's policies stipulate that employees have both the right and the responsibility to apply to internal job openings to advance their career. This policy means that employees do not have to be invited to apply, but are required to be proactive in their application behavior.\footnote{Even though in theory employees are free to apply, it could be the case that in practice many internal applications are solicited. Two facts suggest that this possibility does not represent a major caveat for my analysis. First, it is very rare (2\% of job vacancies) that managers  already have some candidates in mind at the time they post a job and dropping these events (which are recorded due to a firm policy to effectively distribute recruiting assistance) does not alter my results. Second, survey evidence discussed in Section \ref{sec:applications} suggests that unobserved solicitations happen similar to men and women.} 

Quarterly application rates for internal positions are 3\%. Among employees who stay with the firm during my five-year sample period, 27\% submit at least one internal job application. While employees can choose to apply to multiple positions at the same time, the median applicant applies to only one internal position in a given quarter.  The vast majority of applications (97\%) are to positions outside of employees' current team. Similar to the external labor market, 93\%  of applicants have not previously worked with the hiring manager of the position they are applying for. In addition, one-third of internal applications are for positions in a different city.  58\% of internal applicants are invited to an interview and only 28\% of internal applications are successful. Taken together, these patterns suggest that internal applications are typically made under uncertainty.

\section{Data: Personnel Records, Applications, and Survey Responses}
\label{sec:data} 
I assemble a new dataset that combines internal personnel records with application and job vacancy data, allowing me to analyze whether and why men and women make different application decisions. To capture employees' perceptions regarding higher-level positions, I supplement the administrative data with large-scale surveys I conduct with the employees in my sample. 

I collect the firm's internal personnel records from 1998 to 2020, which provide detailed information on demographics and position characteristics for all employees in my sample. The richness of these data allows me to account for key differences between men and women that may influence workers' career progression. 
My data contains detailed demographic information (e.g., gender, age, educational qualifications, family status, and parental leave history at the firm) as well as position characteristics (e.g., position title, function, location, the number of direct reports). I supplement these data with payroll information, capturing employees’ working hours, earnings, and bonus payments. From the firm's talent management system, I also collect information on worker evaluations, such as performance and potential ratings, which are conducted by a worker's direct supervisor. 

To examine employees' application decisions, I collect the universe of application and job vacancy data from 2015 to 2019 at the firm. In order to switch positions at the firm, employees are required to actively apply for an internal job opening through a centralized online portal. From the portal, I collect information on the timing and identity of each application at the firm, capturing both applications from existing employees and from external applicants. In total, the application data cover each of the over 16,000 job openings and around 200,000 applicants from 2015 to 2019. Because I also observe the outcome of each application in terms of rejections, interview callbacks, offers, and subsequent hiring outcomes, I am able to separately measure employees' application choices from the firm's interview and hiring decisions. 

Because the firm requires job openings to be posted to an online job portal, I am able to collect the original posting for the universe of vacancies at the firm from 2015 to 2019. Extracting features from job postings in addition to information from personnel records has several advantages. First, by using the description in the posting, I am able to identify job characteristics that are salient to applicants when they apply. Second, the postings capture relevant job characteristics that are usually not contained in personnel records, and which may likely vary across the leadership ladder, such as required communication skills or assertiveness. Third, in addition to job characteristics, this approach also allows me to control for other relevant features of postings, for instance the language with which jobs are described. To measure how appealing a job seems, I draw on characteristics of the applicant pool. On average, a vacancy receives 4 internal and 23 external applicants. I use the number of external applicants (who were not employed by the firm at the time of application) to measure the relative attractiveness of a job opening conditional on the job's other features. 
 
 For my main analysis sample, I combine the personnel records and the internal application histories into an employee-by-quarter dataset spanning 2015 to 2019. I choose to collapse the data to a quarterly level because the median applicant applies only to one internal position in a given quarter. I restrict my sample to only white-collar and management employees who are regular employees at the firm (e.g., excluding marginal employment such as mini jobs). My main analysis sample contains over 400,000 employee by quarter observations and covers over 30,000 unique white-collar and management employees. 

In order to account for the potential impact of other vacancy characteristics as drivers of the gender leadership gap, I create an auxiliary employee-by-vacancy dataset from 2015 to 2019 that combines each employee in my quarterly analysis sample with every available job opening they could have applied to. I refine these choices based on observed application patterns, dropping combinations that never occur in the data.\footnote{For instance, I drop combinations between employee location and vacancy location for which applications never occur in my data. Due to the large size of the dataset I  restrict to lower-level employees who have applied to at least one position during my sample period. In unreported results, I find similar patterns when using a random sample of employees in lower-level positions rather than restricting to those who have applied at least once.} The dataset includes detailed information about each vacancy's (advertised) job features, information about employees' current positions, as well as their demographics. The dataset also contains an indicator for employees' realized application choices. In total, the dataset has over 39 million observations. In a given quarter, the median employee has 421 job openings in their choice set, yielding over 39 million total observations. 

 To capture employees' perceptions with respect to career progression and leadership responsibility, I design and conduct two large-scale surveys with the employees in my analysis sample. Employees were invited via e-mail by the firm's HR department. Both surveys received a 50\% response rate. Survey respondents are  similar to non-respondents (Appendix Table \ref{table:survey_X}).\footnote{Because of stringent data protection regulations in Germany, I am not able to link survey responses to administrative employee records at the individual level. However, comparing summary statistics between the overall population and survey respondents, I find high similarities both in worker demographics and position characteristics.} I find no evidence for differential selection into response by gender when comparing employees who responded before a reminder was sent to those who responded after (Appendix Table \ref{table:survey_reminder}). The patterns observed in the survey data highly resemble realized outcomes in the administrative data. For instance, while the administrative data suggest that women in lower-level positions are 27.4\% less likely to apply for early-career promotions, the corresponding difference is  33.9\% in the survey. I also find that patterns across both surveys, which were advertised differently, are highly similar. 
 
The first survey asked employees to provide their perspectives on the internal labor market. Survey questions capture employees' perceptions about their career progression at their firm, job search, and the extent and nature of job recommendations received from others. The second survey focused on responsibility over a team and elicited detailed information on perceptions, hypothetical choices, and experiences related to leading a team.  Both surveys include a mix of multiple-choice answers and free-form text responses. The median response time was 13 minutes in the first survey and 16 minutes in the second survey. For my main analysis, I drop incomplete responses.  To probe whether my results are specific to the German context, I conduct a supplementary survey with the firm's white-collar and management employees based in the United States.  This survey used the same question wording as the two surveys in Germany, received a 36\% response rate, and allows me to replicate my main findings for the US setting. Additional details about the survey implementation and an abbreviated version of the survey instrument are provided in Appendix Section  \ref{sec:app_survey}.

\section{Measuring Promotions Along the Leadership Ladder}
\label{sec:hierarchy} 
The goal of this study is to test whether and why climbing the leadership ladder is less appealing to women relative to men. Such an analysis requires a measure that identifies higher-level positions. Classic theories of the firm characterize higher levels of the job hierarchy as exhibiting more authority (\citealp{Rosen1982}). In practice, such increases in authority can take different forms, for instance by taking on responsibility over people, projects, or business decisions (\citealp{SHRM2021}). The employee survey I conduct at the firm corroborates this fact: Respondents indicate that transitions to higher-level positions are characterized by changes along a range of different dimensions, including more project responsibility, leadership of a team, and more autonomy over business decisions (Figure \ref{fig:survey_hierarchy_salient}). 

The ideal measure of hierarchy distinguishes positions based off of their actual leadership characteristics, rather than correlates of leadership like pay. To construct such a measure, I leverage the rich firm data and construct direct measures of job authority. Based on these measures, I classify applications for promotions as those applications that would induce a transition to a higher-level position with more authority.\footnote{Note that at the firm, employees are required to actively apply to make any type of job transition. Promotions in this setting are thus not simply decided by an employee's supervisor, but are job transitions that result in higher-level positions after an employee has applied and got hired for an internal position.} Motivated by evidence across different organizations (\citealp{SHRM2021}), I focus on three relevant dimensions of job authority: direct reports, managerial autonomy, and reporting distance to the CEO.
 
\vspace{3mm}

\noindent \textbf{Direct reports.}--- I measure an individual's authority over other workers by collecting information on the number of workers that directly report to an individual. The number of direct reports is a continuous measure of the extent of authority over a team, for instance with respect to hiring, firing, and performance review.

\vspace{3mm}

\noindent \textbf{Managerial autonomy.}---The firm categorizes each position into five levels of increasing managerial autonomy: (i) positions with neither autonomy over working hours nor decision-making, (ii) positions with autonomy over working hours, (iii) positions with some profit and loss responsibility, (iv) positions with high profit and loss responsibility, and (v) positions at the highest levels, with full responsibility over business decisions. This categorization reflects the extent of managerial autonomy over decisions, which can be independent of authority over direct reports. 

\vspace{3mm}

\noindent \textbf{Reporting distance to the CEO.}---Previous work has noted that reporting relationships can be used to identify where a position falls in the path of decision-making, which captures an important dimension of job authority \citep{Bgh1994}. I collect information on the reporting relationships in the firm by linking employees to their supervisors (and their supervisor's supervisor). This allows me to construct a reporting distance to the CEO for each employee, resulting in eight different levels which resemble the structure of an organizational chart. In contrast to direct reports and managerial autonomy, reporting relationships are periodically adjusted to reflect changes in a firm's focus on specific products or processes. These changes typically do not coincide with employees actually switching positions or taking on different tasks, meaning that this measure of authority is a function of the current organizational structure of the firm.

\vspace{3mm}

While each of these dimensions of job authority can be used separately to identify promotions, my preferred measure of job hierarchy combines all three dimensions into a one-dimensional ranking. This approach is conceptually appealing because it makes comparisons across different areas of the firm more interpretable, which is particularly important in settings where complex internal labor markets are studied. For instance, positions may differ in the number of reports not only because of the extent of responsibility they entail, but also because team size varies across areas of the firm. For example, at the firm teams in engineering tend to be generally larger than teams in finance, raising the question whether the same number of reports across different areas of the firm reflects the same level of hierarchy. By also accounting for the extent of managerial autonomy and the reporting distance to the CEO,  positions can be differentiated from those with similarly many reports that do have more responsibility. This added comparability is especially important given that men and women systematically sort into different areas of the firm (Table \ref{table:sample_descriptive}).  In addition, using different complementary measures to test for potential  differences in the extent to which men and women seek promotions has the advantage that it does not require specifying ex-ante which dimensions of leadership (e.g., over people, over business decisions) are differentially appealing  by gender.

For my preferred approach, I therefore construct a combined measure that incorporates the three dimensions of job authority: number of direct reports, managerial autonomy, and reporting distance to the CEO. Specifically, I use  the first principal component of the three dimensions of job authority, which explains 61\% of the variation and provides a consistent order of all positions at the firm. All three inputs are similarly important and load on the first component as follows: $0.5591\times \{\text{number of reports}\} + 0.6336\times \{\text{managerial autonomy}\}  + 0.5348\times \{\text{reporting distance to the CEO}\}$. The fact that the three dimensions similarly load on the first component indicates that this measure is well-suited to its purpose of combining complementary inputs to enable comparisons across different positions.  The resulting one-dimensional hierarchy  ranking ranges from 0 to 100, assigning the lowest  rankings to entry-level positions (e.g., junior engineering positions), while the position of the CEO receives the highest ranking of 100. Panel A of Figure \ref{fig:hierarchy_index_distribution} shows the distribution of the  hierarchy ranking in my sample. Similar to other hierarchy measures in the literature, I  find evidence for a pyramidal structure, where the vast majority of employees works in lower-level positions. Because in my data a supervisor's ranking is on average 20 points higher than that of her workers, I define promotions as transitions with an increase in ranking of at least 20.  For employees in lower-level positions, the vast majority of early-career promotions defined in this way result in a transition with a meaningful change in job responsibility (e.g.,  promotion from specialist to team leader or from engineer to head of project).

Several robustness exercises corroborate the validity of this measure of hierarchy. First, I find that the hierarchy ranking is strongly correlated with earnings, suggesting that it captures meaningful differences between positions (Appendix Figure	\ref{fig:hierarchy_index_earningscorr}). However, Appendix Figure \ref{fig:hierarchy_index_earningscorr} also shows that the ranking is more effective at discerning between levels at the bottom of the hierarchy relative to pay. This pattern is not surprising, given that in many organizations pay is impacted by a variety of factors and often does not explicitly take nuanced differences in job responsibility into account, especially for lower-level positions. Second, I find that the hierarchy ranking captures important differences in position characteristics. Appendix Table \ref{table:hierarchy_level_x} documents, for instance, that bonus payments represent larger shares of employees' total compensation as hierarchy levels increase, which suggests that employees have more autonomy in higher-level positions.  

Third, increases in the hierarchy ranking correspond to typical steps in the job ladder.  Appendix Table \ref{table:hierarchy_transition} presents a transition matrix for employees who switch positions and shows that employees are most likely to move to adjacent hierarchy levels.  Moreover, Appendix Figure \ref{fig:hierarchy_position_type} shows that hierarchy levels correlate well with markers of seniority in position titles. Even though position titles do not always reflect the type of responsibility a position entails, the share of position titles that reference enhanced job responsibility (e.g., ``team lead'',  ``head of department'') rises as hierarchy levels increase. This finding corroborates the ordering of positions that the hierarchy ranking induces. However,  Appendix Figure \ref{fig:hierarchy_position_type} also illustrates that using position titles alone likely fails to distinguish between leadership levels, as many position titles do not sufficiently reflect how much  responsibility a position entails. For instance, some positions at high hierarchy levels are labeled as engineering or specialist positions, even though they entail leadership over a department. 

My hierarchy measure, which is possible because of the uniquely rich nature of my data, carries important advantages relative to the two common approaches in research relating to internal job hierarchies. The first common approach is to use pay-based measures, such as individual salaries, the salary band, or the job grade a position is assigned to (\citealp{BronsonThoursie2019}, \citealp{CullenPerezTruglia}). This approach reflects increased job responsibilities in higher-level positions to the extent to which they are remunerated differently from tasks in lower-level positions. However, salary differences may also be affected by other factors, such as market wages, candidates' negotiation success, and whether outside offers are matched. If these alternative factors differ by gender, for instance due to gender differences in negotiation (\citealp{Bertrand2011}), this approach may define higher-level positions differently by gender, leading to biased comparisons. In contrast, my use of objective characteristics of job authority is not affected by such bias. Since my data also capture pay, I am able to make comparisons using this existing approach based on pay, which I present along with my preferred measure.

The second common approach is to use flows between different position titles and infer promotions as common transitions from one position title to another. This revealed preference approach is appealing in settings in which internal labor markets are homogeneous and consist of a relatively limited number of positions.\footnote{For instance, the firms studied in \cite{Huitfeldt2021} have an average of ten occupations, and the firms in \cite{Bgh1994} (a medium-sized service-sector firm) and \cite{RansomOaxaca2005} (a supermarket) have a relatively small set of possible career trajectories.} However, this is not the case for many firms, particularly larger firms. In the firm that I study, the internal labor market consists of over 200 different occupations and multiple non-intersecting career paths, making it difficult to construct a universal hierarchy ranking based on transitions between position titles. Another challenge for more complex labor markets is that position titles are often both noisy and relatively coarse, making it difficult to make granular distinctions between different hierarchy levels. For example, in my sample, 26\% of employees share a position title with either their supervisor or their supervisor's supervisor. My use of direct measures of job authority allows me to avoid relying on (noisy) position titles.

\section{Gender Differences in Applications for Promotions}
\label{sec:applications}
While many organizations require employees to actively apply to make internal job switches (\citetalias{hkp}), survey evidence suggests that some features of higher-level positions may be less appealing to women than to men (\citetalias{HBR2014}). This section therefore tests for gender differences in the likelihood of applying for promotions to higher-level positions. 

I estimate gender differences in applications using a logit regression of an indicator for applying in a given quarter on gender and quarter fixed effects. Gender gaps are computed by dividing the average marginal effect for women based on the logit coefficient by the outcome mean.
\vspace{-2mm}
\begin{align*}
	\text{Pr}(\text{Applied}_{it} = 1) = \Lambda(\theta_1 \text{Female}_{i} + \theta_X X_{it}   + \theta_t)
\end{align*}

I leverage the richness of my data to account for underlying differences between men and women that may influence their application decisions. Because of systematic sorting into different areas of the firm by gender (see Table \ref{table:sample_descriptive}), female employees may experience different application opportunities than male employees who work in different areas at the firm.  Prior research has also highlighted the importance of factors such as employees' family status and hours requirements for gender differences in labor market outcomes. In my preferred specification,  I include a broad set of controls $X_{it}$, which capture worker demographics (age, German citizenship, educational qualification, marital status, family status, parental leave, firm tenure) and position characteristics (function, division, location, full-time, weekly hours, responsibility over a team). While the advantage of this specification is that my results can be interpreted as application differences between men and women with similar characteristics who could have applied to similar positions, I document in robustness exercises that my results do not hinge on a specific set of covariates (Appendix Table \ref{table:robust_controls}). 

As discussed in Section \ref{sec:hierarchy}, I identify promotions using detailed information on three relevant dimensions of job authority in the spirit of \citet{Rosen1982}: the number of direct reports, the extent of managerial autonomy, and the reporting distance to the CEO. My preferred approach uses PCA to combine these dimensions into a one-dimensional hierarchy ranking and defines promotions as transitions to positions with a higher ranking. To probe the robustness of my results, I use pay increases as an alternative measure for identifying promotions. Instead of only relying on the combined measure, I also show gender differences in applications when separately using the three dimensions of authority to identify higher-level positions.

My main analysis focuses on employees in lower-level positions. I study applications among these workers in order to focus on the first step along the leadership ladder. Taking this step has long-term career impacts because firms prioritize internal candidates when filling higher-level positions (\citetalias{hkp2023}). In addition, a growing body of evidence documents that important gender differences with respect to career progression emerge early in employees' careers, further motivating a focus on lower parts of the leadership ladder (\citealp{BronsonThoursie2019}, \citealp{Hospido2019}, \citealp{Friebel2022}, \citealp{azmat2020gender}). In line with my preferred approach of identifying higher-level positions using the combined measure of job authority, I define lower-level positions as those with a hierarchy ranking of 20 or less, which captures employees who work as individual contributors without enhanced job responsibility. Appendix Table \ref{table:robust_group} documents robustness to using alternative approaches to identify lower-level positions.

I find that women at lower rungs of the leadership ladder are substantially less likely to apply for promotions relative to their male counterparts. Using my preferred approach for identifying promotions based on the combined measure of job authority, I find a gender application gap of 27.4\% ($p$=0.000, Column 1 of  Table \ref{table:applied_promotions}). Similar patterns result when separately using the measures of job authority to identify higher-level positions. Column 2 shows that when using increases in direct reports as a measure for promotions, women are 30.2\% ($p$=0.000) less likely to apply. Similarly, Column 3 documents an application gap of 15.4\% ($p$=0.000) with respect to promotions defined by increases in managerial autonomy over hours and business decisions. I do not find any meaningful differences when only using the reporting distance to the CEO as a measure for higher-level positions (Column 4).  A key difference between reporting distance and the other measures of authority is that reporting distance is not necessarily an inherent characteristic of the position. An employee's reporting distance can change, for example when a firm shifts its focus on specific products or processes, without any changes to the employee's position, suggesting that it may be a less crucial prerequisite for application decisions. The finding that there are large gender differences by direct reports and managerial autonomy, but not reporting distance to the CEO, motivates a deeper investigation into the features of leadership that appeal differentially by gender, which I return to in Section \ref{sec:teamleadership}.

Using pay increases as an alternative measure of promotions (Column 5 of Table \ref{table:applied_promotions}) also reveals a gender difference in applications of 14.7\% ($p$=0.041). Although pay by itself is a correlate, rather than a direct measure of leadership, this finding supports the overall conclusion that there is a substantial gender gap in applications for higher-level positions. Together, my results suggest that across different measures of job hierarchy, women in lower-level positions are substantially less likely to apply for early-career promotions. In light of the robustness of my findings,  I use my preferred approach based on the combined measure to identify applications for promotions in the remainder of the paper. 

The magnitude of the gender gap in applications is striking given that these estimates control for detailed position characteristics, demographics, and employee qualifications. This finding indicates that women who have a very similar background as their male counterparts and who face similar career opportunities are substantially less likely to apply for promotions. Note that although quarterly application rates are low, the gaps accumulate to affect a much larger share of potential applicants. Specifically, 7\% of men who start out in a lower-level position apply for a promotion during the five years of my sample, whereas women are 23\% less likely to ever apply. In addition, the observed application gap is not explained by differential selection into or out of the sample, but exists also when restricting to the subset of employees who are continuously in my sample during the sample period (Column 5 of Appendix Table \ref{table:robust_controls}).

My findings demonstrate the existence of a broken rung:  women in lower-level positions are substantially less likely to apply for promotions than men. Yet at the same time, women at higher rungs of the leadership ladder are not less likely to apply for subsequent promotions. Table \ref{table:applied_main} uses my preferred hierarchy measure and finds no gender differences in applications among employees who already hold a first-level leadership position---either in the form of leading a team, having responsibility over a project, or working as an expert (Column 2). Similarly, when comparing men and women in any type of leadership position, spanning positions with enhanced job responsibility  up to the CEO, I also find no evidence for meaningful application differences (Column 3 of Table \ref{table:applied_main}). A similar pattern emerges when using alternative measures of job hierarchy. Appendix Table \ref{table:robust_pay} uses pay increases to identify promotions and also finds that women at higher rungs of the job ladder are not less likely to apply for higher-level positions relative to their male counterparts. These results suggest that early-career promotions may be of particular importance for understanding gender differences with respect to leadership. In the remainder of the paper, I therefore focus on employees in lower-level positions in order to investigate why these differences in applications for promotions occur. Appendix Section \ref{sec:appendix_leaders} provides analogous results for employees in leadership positions and documents the absence of meaningful gender differences along a range of key career-related actions and perceptions.

\subsection{Can Worker-Level Factors Explain the Application Gap?}

One potential interpretation of the gender gap in applications for promotions is that it is driven by differences in worker characteristics. For example, prior work analyzing applications for new positions in general (i.e., not promotions specifically) has documented that gender differences in hours constraints (\citealp{MasPallais2017}, \citealp{WiswallZafar2018}), preferences for location  (\citealp{LeBarbanchon2021}), and confidence(\citealp{Coffman2020}) are important determinants of gender disparities in labor supply decisions. Therefore, it is possible that these characteristics can also explain the gender gap in applications to promotions, rather than specific features of promotions that are inherently less appealing to women. In order to assess whether this is the case, I test whether controlling for key worker-level characteristics can account for the observed differences in applications for promotions. 

\vspace{2mm}

\noindent \textbf{Change in work environment.}---The observed gender gap in applications for promotions could arise if having to undergo changes in the work environment, independent of increases in job hierarchy, makes job switches differentially appealing to women and men. When focusing on lateral switches, I find that women are more likely to apply relative to their male counterparts (Column 1 of Table \ref{table:applied_lower_other}), suggesting that it is unlikely that gender differences with respect to \textit{any} changes are a key driver in my setting.  Columns 2 and 3 of Table 	\ref{table:applied_lower_other} further corroborate this finding. Column 2 focuses on applications for lateral switches that require a transition to a different function within the firm, which typically entail crucial changes to employees' job tasks (e.g., moving from engineering to sales). I find that women are not less likely than men to apply to positions in different functions. Column 3 focuses on applications for lateral switches to a different  division of the firm, which involve changes to employees' work environment (e.g., different culture). I find that women are more likely than men to apply for these switches, suggesting that having to change work environments alone is unlikely to explain my results.  In the  employee survey, I also find no meaningful difference in the share of men and women who report wanting to stay in a position that is similar to their current position (Figure \ref{fig:survey_aspiration}).

Another salient change in employees' work environment is the location of a position. If promotions  disproportionally require employees to switch locations, gender differences with respect to commuting may account for the observed application gap  (\citealp{LeBarbanchon2021}).  To test whether my findings are driven by gender differences with respect to job location, I distinguish applications for promotions by whether the new position is based in a different city relative to employees' current workplace. Column 5 of Table 	\ref{table:applied_lower_other} focuses on applications for promotions in the same city, while Column 6 focuses on promotions in a different city. Even when restricting to promotions that would occur in the same location, women are 30.1\% less likely to apply, suggesting that distance  is unlikely to be the key driver of the observed gender gap in applications for promotions.\footnote{Note that even though location alone cannot explain my results, I do find meaningful gender differences in applications if the new position is far away. For instance, women are 16\% less likely to apply for a position that is more than 100 kilometers away, suggesting that the requirement of having to relocate may differentially affect men and women, in line with findings by \citet{LeBarbanchon2021}.} 

\vspace{2mm}

\noindent \textbf{Hours constraints and children.}---Applications for promotions may also differ because women in lower-level positions are less likely to work full-time and thus may not consider applying for any positions that require working full-time, including most leadership positions. I test this possibility by assessing heterogeneity in employee demographics. Column 1 of Table  \ref{table:applied_hetero} shows that even when focusing on the subset of employees who already work full-time and for whom the requirement to work full-time is less binding, I find a gender gap of 34.0\%. Previous literature has also focused on the presence of children as an indicator for a higher demand for flexibility. Even among the  employees who currently do not have children, women are 22.0\% less likely to apply for promotions (Column 2 of Table  \ref{table:applied_hetero}). These findings suggest that even among employees for whom hours constraints are less binding, meaningful gender differences in applications for promotions exist.

A related reason why women may be less likely to apply for promotions is family planning. For instance, if many women in my sample foresee having children in the near future, they may refrain from applying for promotions because of their future family status.  Two pieces of evidence suggest that planned fertility is unlikely to be the sole driver of the observed application gap for promotions. First, I find large gender differences  even among employees who are not married and thus are less likely to start a family soon (Column 3 of Table \ref{table:applied_hetero}). Second, I find in unreported results that large application gaps exist even for employees above the age of 42 for whom future family planning is less likely to be relevant. While my analysis is not set up to test whether long-term family planning differentially affects how men and women make decisions with respect to their education and early career, my results suggest that even men and women who are unlikely to start having children in the near future make very different application decisions with respect to promotions. 

\vspace{2mm}
\noindent \textbf{Search Costs.}---Women may also be less likely to apply for promotions because of the time and effort costs related to finding suitable positions. Previous work has shown that in the external labor market, such costs give rise to differential search outcomes by gender (\citealp{Skandalis2022}). Several pieces of evidence suggest, however, that in the internal labor market I study, gender differences in search and  in the access to information are limited. 

First, the firm has a strict policy in place that requires all jobs to be posted to a centralized job portal, implying that all employees in my sample have easy access to each job opening.  Applications through the job portal take less than five minutes to complete and only require employees to fill in their contact information online, limiting the time effort of applying.  Second,  I do not find that women are less likely to apply for lateral switches (Column 1 of Table \ref{table:applied_lower_other}), even though the time effort related to applying is the same for all types of applications in the internal labor market.

Third, while in theory men and women could differ in their usage of the job portal, responses from the employee survey do not reveal any evidence for realized differences in information. Men and women in lower-level positions are equally likely to report actively searching for jobs (Bars 1 and 2 of Appendix Figure  \ref{fig:survey2019_search}, Panel A). They are also similarly likely to get approached by others with recommendations about internal job openings (Bars 3 and 4 of Appendix Figure	\ref{fig:survey2019_search}, Panel A). In addition, the type of individual who makes these recommendations (e.g., supervisor vs. teammate) is similar for men and women (Appendix Figure \ref{fig:survey2019_rec}). When asked what would be most supportive for their internal career progression, only 4.5\% of male and 4.5\% of female respondents name better access to information about available job openings as important factor, suggesting that men and women are similar regarding the information they receive about the availability of job openings. 

Finally, I find large gender differences in applications for promotions among employees for whom search costs are less binding. Column 6 of Appendix Table \ref{table:robust_controls} shows that among employees who have applied at least once in a given quarter, and who thus have to be aware of the job portal, women are 17.9\% less likely to apply for promotions. In addition, previous work has documented the importance of homophily with respect to information provision, suggesting that search costs could be lower in areas with more women (\citealp{jackson}).  However, Table \ref{table:applied_hetero} shows that large application gaps exist in female-leaning areas (Column 4), under a female supervisor (Column 5), and in a team with an above-median female share (Column 6).

\vspace{2mm}

\noindent \textbf{Application success.}---Another potential explanation for the observed gender differences in applications is that women may perceive a lower success likelihood conditional on applying. Under such a scenario, it would be the (perceived) hiring process, not the appealingness of higher-level positions, that deters female applicants. Several exercises  suggest that differences in success likelihood are an unlikely driver of the application gap I document.

To the extent to which women base their perception of their own hiring likelihood on their observation of other women's experiences, testing for gender differences in realized success probabilities provides a first piece of evidence for the role of application success. Appendix Table \ref{table:hiring} shows that conditional on applying for a promotion, women in lower-level positions are 23.4\% more likely to be interviewed (Column 1) and 45.1\% more likely to be hired (Column 2) relative to the outcome mean. If women use observed hiring probabilities as a proxy for their own success likelihood, they may factor in potential positive selection into applying. However, given that the gender gap in observed hiring likelihood is large and positive (+45.1\%) and the  gender gap in applications is large and negative (-27.4\%), the extent of negative selection into applying would have to be very large to rationalize that women do not apply because they downward adjust their own hiring probability. 

In presence of information frictions, actual hiring probabilities may not be of first-order importance for employees' perceptions of success likelihood. For instance, women may not apply for promotions because they are worried about gender discrimination in hiring. To match the patterns in the data, this worry would have to be specific to applications for promotions and not for lateral switches. In line with previous research that highlights that the gender of decision-makers is a useful proxy for differential treatment by gender (\citealp{CullenPerezTruglia}), women may be particularly prone to perceive discrimination in male-dominated areas and under a male supervisor.  To account for this possibility, I test whether gender differences in applications arise even in female-leaning areas, which are less likely to be associated with discrimination against women. Columns 4 to 6 of Table  \ref{table:applied_hetero} show that large gender differences in applications exist even in areas with high shares of female peers and under female supervisors.\footnote{Since hiring decisions are made by the supervisor of the job vacancy and not the current supervisor, Table \ref{table:refpref_alternatives} presents results  that control for the gender composition of the vacancies employees could apply to. If find large gender differences in applications even if the vacancy has a female supervisor or has a lot of female coworkers.}  

I corroborate this finding by directly testing whether women and men report perceiving differential success likelihoods in the employee survey.  Respondents were asked to provide advice regarding an employee in a lower-level position who is considering applying for a promotion in form of a team leadership position. The name of the employee was randomly chosen to be either \textit{Matthias} to signal that the employee is male or \textit{Sophie} to signal that the employee is female. Respondents were asked how high the likelihood is that the employee's application is successful on a range from 0\% to 100\%. Panel A of Figure \ref{fig:survey_vignette} shows the share of employees who indicated a high success likelihood defined as a hiring probability of 60\% or more, which corresponds to an above-median response. I find no evidence that women perceive lower hiring likelihoods for female applicants relative to male applicants. If anything, women's  perceptions regarding either type of applicant are slightly higher than men's. 

Even if men and women face similar hiring probabilities, failure may have differential effects on them (\citealp{Buser2016}). I test whether the fear of rejection can explain my results by focusing on the subset of employees who re-apply for an internal job opening after being rejected for a previous application. Because these employees re-apply,  the fear of rejection is likely less binding for them. However, I find that even in this subgroup, women are 26.2\% less likely to apply for promotions (Column 7 of Appendix Table \ref{table:robust_controls}). This channel can also not explain why the gender application gap is limited to promotions.\footnote{Another fear that may differ by gender is that of retaliation, for instance in the context of managerial talent hoarding, which may deter more women from applying than men (\citealp{Haegele2023}). In unreported results, I find a substantial gender gap in applications even when I focus on employees who work under a manager with a low propensity  to hoard talent, suggesting that for them the fear of retaliation is less binding.} 

\vspace{2mm}
\noindent \textbf{Risk, competition, and confidence.}---Previous literature suggests that gender differences in applications could also be predominately driven by differences in risk preferences, willingness to compete, or confidence  (\citealp{Bertrand2011}).  To test for this possibility, I turn to the employee survey, which documents a gender gap in applications for promotions of 33.9\% (Column 1 of Appendix Table \ref{table:applied_pref}). To measure risk preferences and willingness to compete, I use standard survey questions from the literature based on \cite{dohmen2011} and  \cite{buser2021}, respectively.  To measure confidence, I follow  \cite{buser2021} and use a survey question from \cite{rosenberg2015}.\footnote{See Appendix Section \ref{sec:app_survey} for the exact question wording.} In line with  past work, I find that relative to men, women in my sample are more averse to risk, are less willing to compete, and are less confident. However, these differences can only account for a small share of the gap in applications for promotions. Appendix Table \ref{table:applied_pref} shows that even when focusing on employees with high preferences for risk (Column 2), employees with a high willingness to compete (Column 3), and employees with a high confidence (Column 4), large gender differences in applications persist.

\vspace{2mm}

\noindent \textbf{Substitution with external promotions.}---If internal promotions are less appealing to women than to men, women may substitute applications for internal positions with applications for positions outside the firm. Such a behavior may explain the gender differences in applications for internal promotions that I document. While I do not observe applications to other firms, my survey elicited whether women are more likely to search outside the firm.  Panel B of Appendix Figure \ref{fig:survey2019_search} documents the opposite pattern. Women are 10\% less likely to search for positions outside the firm and are 23\% less likely to receive recommendations with respect to external job openings. Women in lower-level positions are also 20\% less likely to exit the firm (Column 2 of Appendix Table \ref{table:entryexit}). This evidence suggests that women are not more likely to substitute internal applications with external ones. 

\vspace{4mm}

\noindent  Collectively, the findings in this section document that the substantial gender differences in applications for promotions I document are not fully accounted for by worker-level factors that previous literature has pointed out. This raises the question why promotions may be less appealing to women than men, which is the focus of the subsequent section. 

\section{The Role of Team Leadership}
\label{sec:teamleadership}
Why are promotions less appealing to women relative to men? Previous research in economics has highlighted that differences in job characteristics, such as  flexibility or commutes, play an important role for gender differences in labor supply (\citealp{MasPallais2017}, \citealp{WiswallZafar2018}, \citealp{LeBarbanchon2021}). However, little is known about other changes in job characteristics that accompany transitions to higher-level positions. Instead, promotions are typically measured by increases in pay or flows between occupations, prohibiting conclusions about  what it is about a higher-level position that may be less appealing to women than to men.  

\begin{figure}[!ht]
	\thisfloatpagestyle{plainlower}
	\caption{Changes that Accompany Early-Career Promotions}
	\centering
		\includegraphics[scale=1.0]{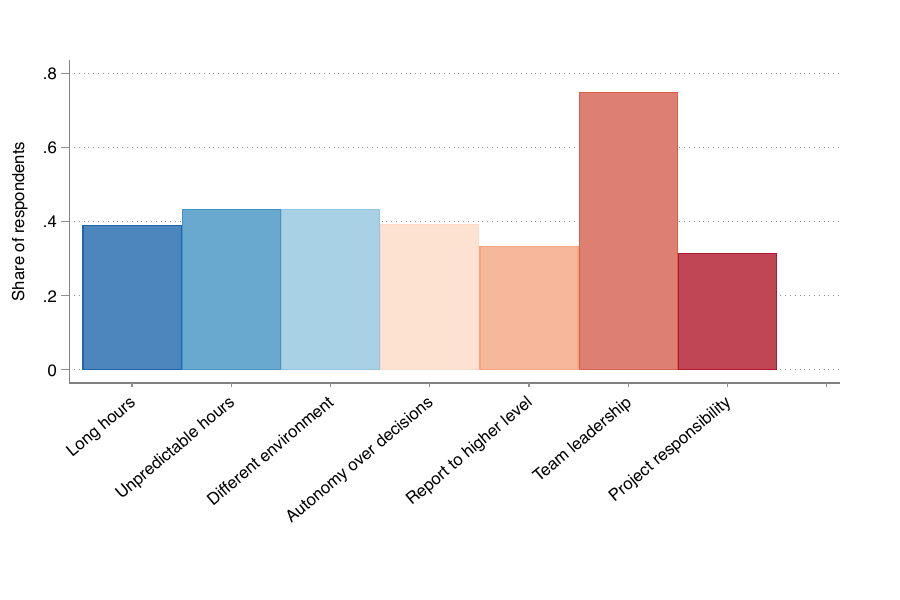}
	\label{fig:survey_hierarchy_salient}
	\begin{minipage}[b]{0.8\linewidth}
		\footnotesize \textit{Notes}: This figure shows employees' perceptions of which changes accompany early-career promotions based on the employee survey. Employees in leadership positions were asked to think back to their promotion from a lower-level position and indicate all changes that characterized their transition. Employees could choose among following answers: increase in regular hours or business travel, increase in unpredictable overtime, change in work environment (e.g., different co-workers), more autonomy over critical business decisions, reporting to employees at higher levels, more responsibility because of leadership over a team, more project responsibility. See Appendix Section \ref{sec:app_survey} for the exact question wording.  N$<$5,xxx
	\end{minipage}	
\end{figure}

Responses to the employee survey illustrate that early-career promotions are characterized by  several changes in job characteristics. Employees in higher-level positions at the firm were asked to think back to their first promotion from a lower-level position and indicate which changes they perceived as most salient. Figure \ref{fig:survey_hierarchy_salient} shows that employees name a range of factors, which are in increasing frequency:  greater responsibility over projects (32\%),  a shorter reporting distance to the CEO (34\%), longer working hours (39\%),   enhanced responsibility over critical business decisions (39\%), changes in the work environment (43\%),  less predictable hours (43\%),  and  taking on responsibility over a team (75\%). Having to take on responsibility over a team is named almost twice as often as any other factor, suggesting that this is a very salient dimension of early-career promotions.

While a growing body of experimental research suggests different dimensions of leadership that may give rise to gender differences (\citealp{ertac}, \citealp{ChakrabortySerra}, \citealp{eckel}), the extent to which responsibility over a team is less appealing to women and its role for early-career promotions hasn't received much attention, especially in real-world work settings. In most organizations, however, taking on responsibility over a team is a key requirement for career advancements (\citetalias{hkp2023}), highlighting the importance to study whether men and women differ with respect to the requirement to take on responsibility over a team.\footnote{While not all promotions from lower-level positions require taking on responsibility over a team, 85\% of executive positions and 97\% of top management positions require team responsibility, indicating that it is a key prerequisite for ultimately advancing in the leadership hierarchy.} Empirically detecting the impact of team leadership on gender differences in applications is difficult. Standard datasets contain little information on the extent to which the position involves responsibility over a team. In addition,  responsibility over a team may be correlated with other job characteristics, such as pay or a higher share of male coworkers, which may independently affect women's application likelihood.

\subsection{Direct Evidence from Employee Surveys} 
To isolate whether and why responsibility over a team may be less appealing to women, I designed a new survey instrument that captures how employees perceive team leadership. To elicit direct evidence on the extent to which taking on responsibility over a team is less appealing to women than to men, I use three distinct survey questions. The survey was fielded to all employees in my sample and received a 50\% response rate. Section \ref{sec:data} provides information on the implementation and reveals no indication of meaningful differential selection into response. 

My primary survey question asked employees where they would like to see themselves in five years with respect to their career. Figure \ref{fig:survey_aspiration} presents answers separately for men and women  in lower-level positions. The most frequent answer for men  is to have more pay (26\%), followed by taking on responsibility over a team (24\%), working in a similar position relative to their current one (22\%), more challenges (13\%), more responsibility over a project (10\%), and more flexibility (6\%). Women, however, are 36\% less likely ($p$=0.000) to report wanting to lead a team compared to the outcome mean. Other job characteristics, such as pay or challenges at work are not characterized by any meaningful gender differences.\footnote{While women are more likely than men to desire flexibility, the absolute share of women who does is small (8\%).} Employees who want responsibility over a team are  twice as likely to have applied for an internal position with responsibility over a team in the past 12 months, indicating that differences in where employees would like to see themselves translate into career-relevant actions.  

The observed gender differences with respect to team responsibility are present across different groups of employees at the firm. Column 1 of Table \ref{table:survey_robustaspiration} documents that among employees who work full-time, women are 29\% less likely to see themselves in positions with team leadership. Among employees who do not have children the gender gap is 33\% (Column 2). Even among employees who work in female-leaning area, which are characterized by high shares of female supervisors and co-workers, the gender gap is 32\% (Column 3). These findings echo the robust gender differences in applications for promotions documented in Section \ref{sec:applications}. 

The gender leadership gap is also not accounted for by differences in risk preferences, willingness to compete, or confidence. I use the standard survey questions from \cite{dohmen2011} and \cite{buser2021} to measure risk preferences and willingness to compete, respectively. To measure confidence, I  use a survey question from \cite{rosenberg2015}.  Table \ref{table:survey_robustaspiration} shows that even when I focus on employees with high risk preferences (Column 4), high willingness to compete (Column 5), or high confidence (Column 6), women are at least 30\% less likely to want team leadership.\footnote{In unreported results, I find that even when jointly controlling for my continuous measures of risk preferences, willingness to compete, and confidence, the gender leadership gap is reduced by less than a quarter.} I also elicit whether employees think they will be  effective leaders (see Appendix Section \ref{sec:app_survey} for question wording). However, even among the set of employees who think they will be effective leaders, women are 29\% less likely to want to take on team leadership (Column 6). 

To further corroborate the observed leadership gap, a second survey question asks employees to choose among three types of workshops that the firm considers offering:  a workshop about how to become a successful team leader, a workshop about how to create a career without  responsibility over a team, and a workshop about how to navigate work-life balance. Employees could also indicate whether they preferred none of these workshops. The advantage of this question is that employees were aware that workshops (some of which were actually later provided by the firm) that received little interest may not be offered, suggesting that respondents have an incentive to provide truthful answers to this question.\footnote{Employees were reminded at the beginning of the survey that the firm has used survey responses in the past to implement high-stakes changes. The survey prompted unsolicited emails from respondents who wanted to provide additional feedback, further indicating that the wording made employees feel that their responses were consequential.} 

\begin{figure}[!ht]
	\thisfloatpagestyle{plainlower}
	\caption{Gender Differences With Respect to Team Leadership Workshop}
	\begin{minipage}[b]{1.0\linewidth}
		\centering
		\includegraphics[scale=3.0]{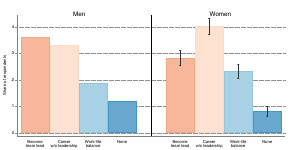}
	\end{minipage}	\\
		\vspace{0.5cm}
	\label{fig:survey_workshop}
	\begin{minipage}[b]{1.0\linewidth}
		\footnotesize \textit{Notes}: This figure uses a complementary question to show gender differences with respect to taking on responsibility over a team based on the employee survey.  Employees were asked which workshop they would chose among  following options that the firm considers to offer at their location: a workshop about how to become a successful team leader, a workshop about how to create a career without leadership over a team,  and a workshop about how to navigate work-life balance. Employees could also indicate if they were interested in none of the workshops. Employees were informed that that the firm is indeed planning to offer one of these workshops and their responses will be important for planning purposes. See Appendix Section \ref{sec:app_survey} for the exact question wording.  This figure restrict to employees in lower-level positions. For men, sample means are shown. For women, outcome means plus the female coefficient are shown.  Controls: Age, tenure, family status, hours, function.  N>5,xxx
		\end{minipage}	
\end{figure}

Figure \ref{fig:survey_workshop} documents that the most frequent workshop choice among men in lower-level positions is the workshop on how to become a successful team leader (36\%), followed by the workshop on creating a career without responsibility over a team (33\%), and the workshop on work-life balance (19\%). 12\% of men are not interested in any workshop. Women, however, are 23\% less likely ($p$=0.000) to choose the workshop on team leadership. Instead, women are 21\%  more likely ($p$=0.000)  to choose the workshop about a career without responsibility over a team and 22\%  more likely ($p$=0.001) to choose the workshop on work-life balance. Women are 35\% less likely ($p$=0.000) to indicate that they do not want to participate in a workshop. These results corroborate the finding from my primary survey question that taking on responsibility over a team  is less appealing to women than to men, but that women are not generally less interested in career development in general.

Because men and women may have different perceptions of what responsibility over a team entails when answering the survey, I designed a third question that poses a hypothetical job choice between positions with and without responsibility over a team, while holding other job characteristics constant. Employees were asked to choose between a position that is similar to their current position, but includes responsibility over a small team, and a position that is similar to their current position, but does not involve responsibility over a team. This definition was designed to make employees' perceptions of the position as similar as possible. Indeed, I find that employees who would like to lead a small team are 143\% more likely to have applied for an internal position with responsibility over a team in the past 12 months, suggesting that this hypothetical choice question likely captures differences among employees that are related to their high-stakes application decisions. However, women are 13\% ($p$=0.000) less likely to choose responsibility over a team, even if the team would be small and all other job characteristics are held constant. 

Together, my findings provide robust evidence in support of the gender leadership gap.  The responses to the three different survey questions yield similar qualitative and quantitative conclusions. I also find that the gender leadership gap is similar in magnitude across different survey waves at the firm that were advertised differently, suggesting that differential selection into response is unlikely a key driver of this result. While the team leadership gap in the second survey wave is 36\%, it was 32\% in the first survey wave in Germany. Similarly, I find a gender leadership gap of 39\% among the firm's employees based in the United States (Appendix Section \ref{sec:appendix_us}). Overall, these findings provide direct evidence that taking on responsibility over a team is less appealing to women than to men.  

\subsection{Complementary Evidence from Realized Applications}
\label{sec:teamleadership_vacancy}


 To corroborate whether team leadership indeed deters applications, I leverage the fact that not all promotions require taking on responsibility over a team and compare realized applications for team leader positions to applications for other types of higher-level positions. The advantage of this complementary approach is that it identifies gender differences based on employees' high-stakes application decisions and allows me to control for an unusually rich set of vacancy characteristics.  While it may be possible that team leadership positions differ from other types of higher-level positions in factors that are unobserved in my data, implying that I cannot fully rule out that observed application differences may be influenced by factors other than team leadership, this approach provides complementary evidence to the direct evidence from employees' survey responses.

This section uses data at the employee-by-vacancy level that contain all potential application choices employees could have made in a given quarter, as well as an indicator for their realized applications. See Section \ref{sec:data}  for more information on the dataset. Because not all promotions require taking on responsibility over a team,\footnote{In my sample, 79\% of positions at the first leadership level require leading a team, while the remainder of promotions are characterized by becoming the head of a project or landing an expert role, which come with enhanced job responsibility but do not involve team leadership.}   I estimate a linear probability model that includes an indicator for whether a vacancy requires team leadership and its interaction with gender. Standard errors are clustered at the employee level and applications for promotions are scaled by 100.  In addition to including the set of employee demographics and current job characteristics ($X_{it}$), I use varying sets of vacancy characteristics  $W_{j}$ as controls. 
\begin{align*}
	\text{Pr}(\text{Applied}_{ijt} = 1) &= \delta_1 \text{Female}_{i} + \delta_2 \text{Team Responsibility}_{j} + \delta_3 (\text{Female}_{i} \times \text{Team Responsibility}_{j}) \\ &+  \delta_X X_{it}  + \delta_W W_{j}  + \delta_t 
\end{align*}

My findings corroborate the importance of team leadership for the gender gap in applications for promotions. Column 1 of Table \ref{table:refpref_top} documents that men are more than twice as likely to apply for a promotion if it requires taking on responsibility over a team, but this is not true for women, yielding a large application gap for promotions that involve responsibility over a team. This finding suggests that a key reason why early-career promotions may be less appealing to women than to men is the fact that the majority of promotions require individuals to take on responsibility over a team for the first time. 

A caveat for this interpretation is that team leader positions may be different from other promotions. To probe whether standard vacancy characteristics may explain my finding, I begin by including a set of controls for the division, function, and location of the vacancy (Column 2 of Table \ref{table:refpref_top}). Column 3 extends the set of vacancy controls to also include key features of the job posting, such as required skills and the length of the posting.\footnote{The set of additional vacancy characteristics includes required work experience, required degree,  required skills (negotiation, problem-solving, communication, analytical, intercultural, assertiveness, English proficiency), character length of the job posting,  and whether the job posting was available in English.} Even when controlling for the increasingly detailed sets of vacancy characteristics, I find that women are substantially less likely to apply for promotions if they involve responsibility over a team.

Given that vacancy characteristics may  affect women and men differentially, I test whether including  interactions between employee gender and a demeaned vacancy characteristic reduces the observed gender gap. For this exercise, I use my baseline set of vacancy characteristics and focus on three potential gender differences that are motivated by previous literature. First, prior work has suggested that women may react differently to skill requirements (\citealp{AbrahamStein}, \citealp{Coffman2020}). If promotions that require responsibility over a team differ also in advertised requirements, gender differences with respect to these requirements may explain my findings.   Table \ref{table:refpref_alternatives} shows results from controlling for differential effects of two key requirements: the requirement to have many years of work experience (Column 1) and the requirement to have more than a Bachelor's degree (Column 2). I find that while these requirements differentially affect men's and women's application likelihood, they do not substantially reduce the gender leadership gap.  

Another dimension in which positions with responsibility over a team may differ is the work environment they impose. Previous work has highlighted  the importance of the gender composition of peers and supervisors (\citealp{CullenPerezTruglia}, \citealp{KunzeMiller2017}).  Table \ref{table:refpref_alternatives} adds controls for whether the vacancy will be in a female-leaning area that is characterized by a high share of female peers and subordinates (Column 3) and under a female supervisor (Column 4). I find that even when controlling for the gender composition of the vacancy's work environment, the gender gap with respect to team leadership remains largely unchanged. 

Finally, promotions that require responsibility over a team may also differ in how they are advertised and consequently perceived by candidates (\citealp{Delfino}, \citealp{Gee2019}). To capture how attractive a job posting appears, I use information from the pool of external applicants. Specifically, I create an indicator for whether the number of external applicants is in the top quartile of all vacancies in my sample. Column 5 of Table \ref{table:refpref_alternatives} shows that even though vacancies with a large external applicant pool are more appealing to employees, this cannot explain the  estimated gender gap in team leadership. In unreported results, I find that the gender gap with respect to responsibility over a team exists even when including all interactions in the same regression. 

Taken together, these results corroborate that team leadership is an important factor for interpreting gender differences in applications for promotions, and that the observed relationship between application behavior and responsibility over a team is not fully accounted for by standard characteristics of the position or the work environment that are likely correlated with responsibility over a team. Because the vast majority of top management positions in organizations require responsibility over a team (\citetalias{hkp2023}), one interpretation of this finding is that men and women differ in their long-term career aspirations and therefore differentially sort into the leadership pipeline (by choosing whether to first take on team leadership). While my study is not set up to distinguish short-term and long-term career aspirations, my findings clearly show that it is not increased job responsibility itself that deters female applicants, but that promotions that come with increased responsibility over projects or other business decisions are indeed appealing to women. 

\section{Implications for Design of Team Leadership Positions}
\label{sec:consequences}
The previous section demonstrated that the requirement to assume responsibility over a team is a key reason why women are less likely to pursue higher-level positions. This finding naturally raises the question of why team leadership is less appealing to women, and how organizations seeking to increase female representation in leadership positions can do so.

\begin{figure}[!ht]
	\thisfloatpagestyle{plainlower}
	\caption{Importance of Position Features of Team Leadership}
		\centering
		\begin{minipage}[b]{0.8\linewidth}
		\centering
		\caption*{Panel A. Reason Why Not Applied for Team Leadership Position}
		\vspace{-4mm}
		\centering
		\includegraphics[scale=0.85]{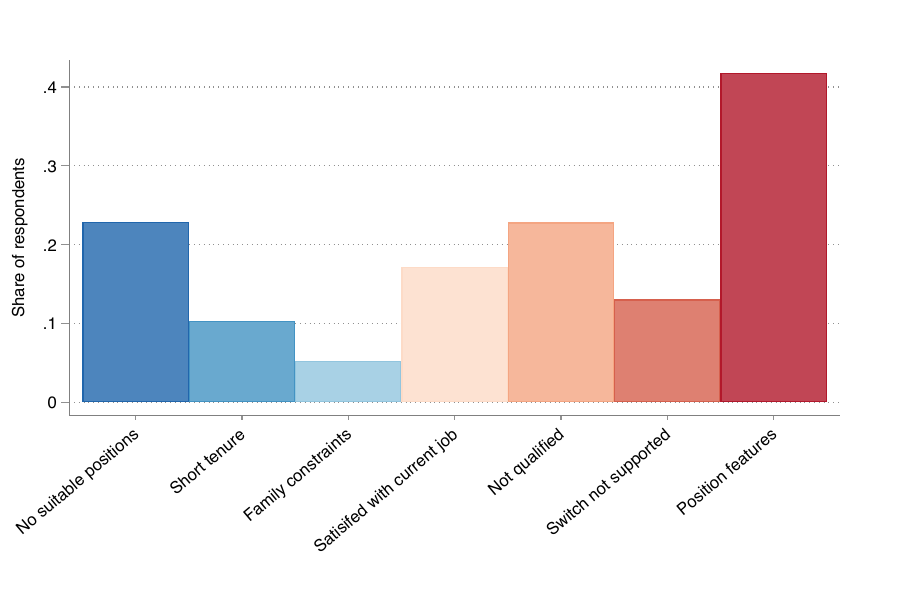}
	\end{minipage} \\	
	\begin{minipage}[b]{0.8\linewidth}
		\centering
		\caption*{Panel B. Hypothetical Job Choice}
		\vspace{-4mm}
		\centering
	\includegraphics[scale=2.0]{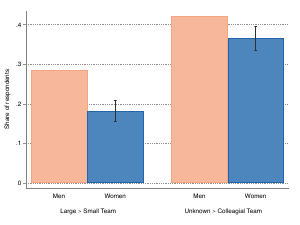}
	\end{minipage}
	\label{fig:survey_hypothetical}
	\begin{minipage}[b]{0.9\linewidth}
		\footnotesize \textit{Notes}: This figure highlights the importance of how team leadership is designed  based on the employee survey. Panel A documents the share of employees in lower-level positions who stated a given reason for why they did not apply for a team leadership position. Free-form responses were categorized into following reasons: no relevant positions  available, short tenure at firm, family constraints,  satisfaction with current position,  not feeling qualified for a team leadership position, switch not supported,  did not apply because of specific features of team leadership. The sample restricts to the subset of employees who did not apply in past 12 months and gave a valid free-form text response, N>4,xxx.  
	Panel B focuses on two key features of team leadership that were listed most frequently as unappealing: large team sizes and the potential of team conflict. Employees in lower-level positions were asked to choose among pairs of hypothetical job openings. Bars 1 and 2 show the share of respondents who choose a job that pays 30\% more and requires leading a large team over a job that pays 10\% more and requires leading a small team, separately by gender. Bars 3 and 4  show the share of respondents who chose a job that pays 30\% more and requires leading a team that they do not know over a job that pays 10\% more and requires leading a team that is known to be colleagial.  See Appendix Section \ref{sec:app_survey} for the exact question wording. Controls: Age, tenure, family status, hours, function.  N>5,xxx
	\end{minipage}	
\end{figure}

 Anecdotal and survey evidence across different organizations suggests that having to lead others is generally negatively perceived by many employees. In a survey of US workers, 66\% of workers report not wanting to lead others (\citetalias{HBR2014}). Reports from HR practitioners and in online worker forums indicate that many workers perceive that positions with team responsibility are overly burdensome, for instance, because they entail responsibility over many administrative tasks, dealing with teammates' personal problems, and conveying negative news to subordinates (\citealp{Inhersight}). Responses from employees in my sample corroborate these patterns. When asked why they did not apply for a team leadership position, 42\% of respondents named the inherent position features of team leadership as the key reason (Panel A of Figure \ref{fig:survey_hypothetical}). This reason was named twice as often as any other, such as not feeling qualified (23\%) and not finding suitable positions (23\%). 
 
 Employees' survey responses point to the role of two position features of team leadership that are perceived as particularly unappealing. First, 55\% of respondents name the administrative burden that team leadership entails, especially for large teams. Employees' explanations from the survey for why this is unappealing include ``Admin and bureaucracy which leaves little time for operative tasks.'' and ``Having to spend all your time on HR tasks, which is a necessity if team sizes are very large.'' Second, 28\% 
 of employees rate the responsibility over resolving conflicts between team members or between the team and upper management as the least appealing feature. Stated explanations include ``Conversations with difficult employees; team leadership positions are often lonely'' and ``Being caught between upper management and your team and having to constantly disappoint your own people.'' See Appendix Table 	\ref{table:survey_quotes} for more quotations from the employee survey that describe why leading a team is perceived as unappealing.

Motivated by these survey responses, I test to what extent the typical design of team leadership positions affects men and women differently based on the two frequently named features of team leadership: the administrative burden as proxied by team size and the potential for conflict.\footnote{In addition to these two features of team leadership positions, another related feature that is named by 5\% of respondents is the constant availability that is required from team leaders. The extent to which team leaders have to be available likely also scales with the size of the team. In light of the importance of hours for potential gender differences, Appendix Section \ref{sec:appendix_availability} provides complementary results analyzing the importance of availability requirements.} To identify potential gender differences, I analyze three possible reasons why these features may affect men's and women's applications for leadership positions differently. First, these features may be less appealing to women relative to men, for instance, because of inherent preferences or associated costs (e.g., conflict aversion). Second, even if men and women do not differ in how appealing they find a given feature, men and women may have different experiences in the same positions. For instance, team leadership may entail more interpersonal conflict for female leaders than for male leaders. Third, even if employees do not perceive differential treatment of leaders, men and women may  differ in their perception of what team leadership entails, regardless of the gender of the leader. Since most leaders are men, women may be less likely to learn about the features of leadership relative to their male counterparts. Resulting misperceptions may deter women from applying, independent of their preferences or costs. I use the survey to assess the relevance of each of these three explanations.
 
\begin{figure}[!ht]
	\thisfloatpagestyle{plainlower}
	\caption{Gender Differences With Respect to Taking on Team Responsibility}
	\centering
	\begin{minipage}[b]{0.8\linewidth}
		\centering
		\includegraphics[scale=3.0]{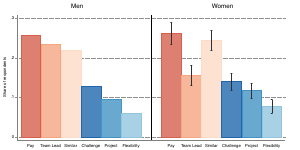}
	\end{minipage}	\\
 		\vspace{0.5cm}
	\label{fig:survey_aspiration}
	\begin{minipage}[b]{1.0\linewidth}
		\footnotesize \textit{Notes}: This figure shows gender differences with respect to taking on responsibility over a team based on the employee survey.  Employees were asked where they would like to see themselves with respect to their career in five years and could choose among following answers:  more pay,  responsibility over a team, staying in a similar job, more challenges, more project responsibility, and more job flexibility.  See Appendix Section \ref{sec:app_survey} for the exact question wording.  This figure restrict to employees in lower-level positions. For men, sample means are shown. For women, outcome means plus the female coefficient are shown.  Controls: Age, tenure, family status, hours, function.  N>5,xxx
		\end{minipage}	
\end{figure}

First, survey responses indicate gender differences in the appeal of the position features of team leadership. Respondents were prompted with a hypothetical job choice question that asked them to trade off higher pay with larger team size and more potential for conflict.\footnote{Note that I find no gender difference in men and women's stated preferences for higher pay (Figure \ref{fig:survey_aspiration}), suggesting that gender differences in the importance of pay are unlikely to confound the results of this exercise.} 
When asked to choose between a position that pays 10\% more than their current position and involves leading a small team and a position that pays 30\% more and involves leading a large team, women are 39\% ($p$=0.000) less likely to choose the higher-paying position that involves leading a large team (Bars 1 and 2 of Figure \ref{fig:survey_hypothetical}, Panel B). Women are also 14\% ($p$=0.000) less likely to choose the higher-paying position that involves leading an unknown team over a team that is known to be collegial, which proxies for a lower likelihood of encountering conflict (Bars 3 and 4 of Figure \ref{fig:survey_hypothetical}, Panel B).\footnote{The advantage of this wording is that is more concrete than asking respondents about trading off the likelihood of conflict. The wording is directly motivated by the fact that survey respondents frequently indicate that not knowing the composition of and atmosphere in the team is a deterring factor for applications because it creates uncertainty about the amount of negative situations team leaders have to manage.} These results suggest  that women are much more likely to avoid the negatively perceived position features of team leadership.

\begin{figure}[!ht]
	\thisfloatpagestyle{plainlower}
	\caption{Employee Perceptions of Differential Treatment}
 \vspace{0.2cm}
	\centering
	\begin{minipage}[b]{0.8\linewidth}
		\centering
		\caption*{Panel A. High Hiring Likelihood}
		\vspace{-2mm}
		\includegraphics[scale=0.7]{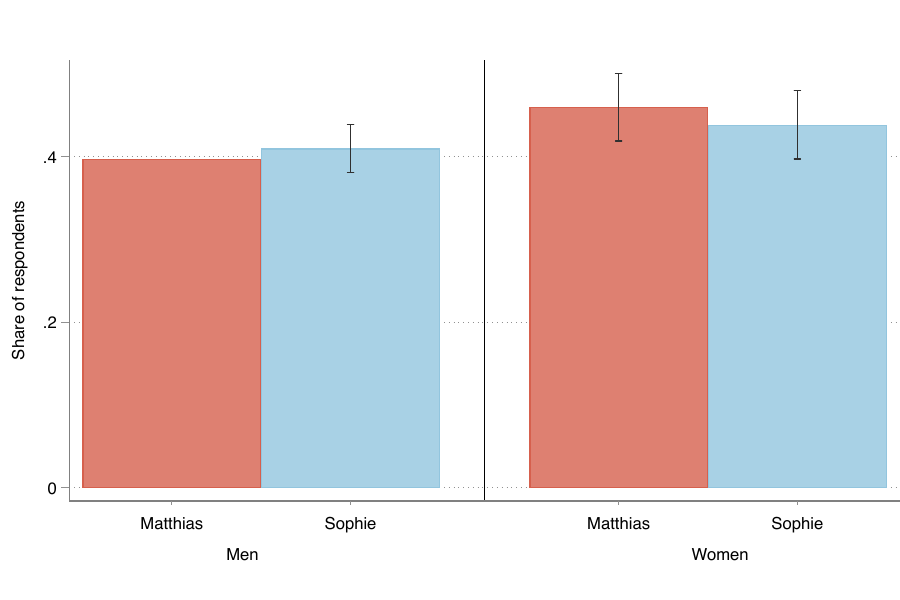}
	\end{minipage}	\\
	\vspace{1.2cm}
	\begin{minipage}[b]{0.8\linewidth}
		\centering
		\caption*{Panel B. Leader Will Have A Difficult Experience}
		\vspace{-2mm}
		\includegraphics[scale=0.7]{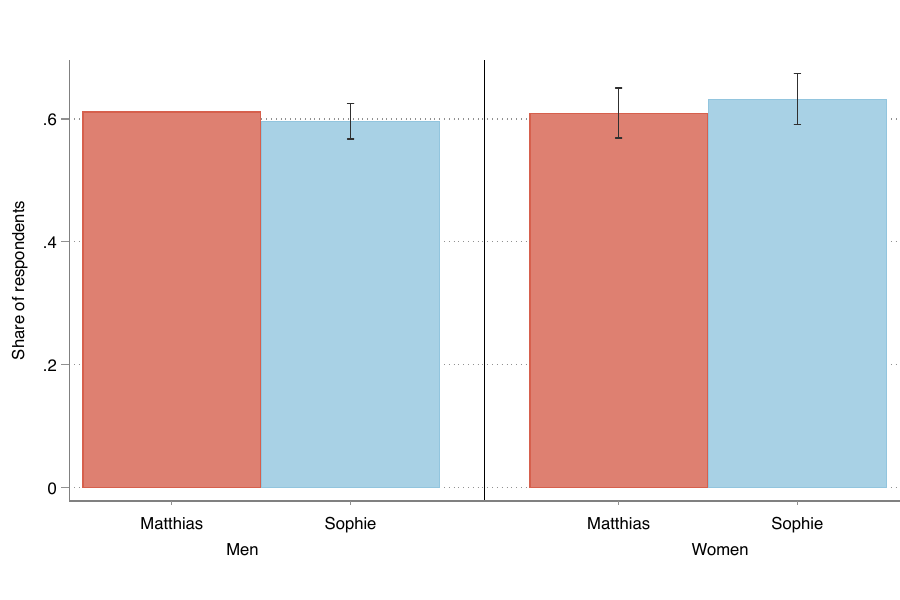}
	\end{minipage}	
	\label{fig:survey_vignette}
	\begin{minipage}[b]{0.8\linewidth}
		\footnotesize \textit{Notes}: This figure illustrates employee perceptions about differential treatment  using a vignette design based on the employee survey. The vignette introduced a fictional colleague whose first name was randomized to be either \textit{Matthias} or \textit{Sophie} to reflect a male or a female gender, respectively. Employees in lower-level positions were asked to indicate how likely an employee would be to get hired if they were to apply for a team leadership position (Panel A) and whether they think they would have a good experience as a leader, especially with respect to  encountering conflict with team members or other leaders if they took on their first team leadership position (Panel B). Panel A shows the share of employees who report a high hiring probability, defined as 60\% or more. Panel B shows the share of employees who report expecting that leaders likely will have a difficult experience.  See Appendix Section \ref{sec:app_survey} for the exact question wording. Controls: Age, tenure, family status, hours, function.  N>5,xxx
	\end{minipage}	
\end{figure}
\begin{figure}[!ht]
	\thisfloatpagestyle{plainlower}
	\caption{Perceptions of Features of Team Leadership}
		\centering
		\begin{minipage}[b]{0.8\linewidth}
		\centering
		\caption*{Panel A. Responses by Current Team Leaders}
		\vspace{-4mm}
		\centering
		\includegraphics[scale=2.0]{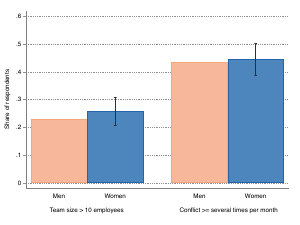}
	\end{minipage} \\	
	\begin{minipage}[b]{0.8\linewidth}
		\centering
		\caption*{Panel B. Responses by Employees in Lower-Level Positions}
		\vspace{-4mm}
		\centering
	\includegraphics[scale=2.0]{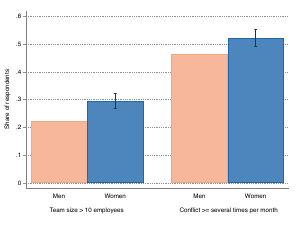}
	\end{minipage}
	\label{fig:survey_perception}
	\begin{minipage}[b]{0.9\linewidth}
		\footnotesize \textit{Notes}:  This figure illustrates gender differences in employee perceptions of team leadership positions based on the employee survey.  Panel A uses responses by current team leaders as benchmark (N<3,xxx). Panel B shows perceptions by employees in lower-level positions (N>5,xxx). Bars 1 and 2 show the share of respondents who report team sizes above 10 employees (median response of current team leaders: 6-10 employees).   Bars 3 and 4 show the share of respondents who report conflict arising at least several times per month year (median response of current team leaders: experience conflict several times a year). See Appendix Section \ref{sec:app_survey} for the exact question wording. Controls: Age, tenure, family status, hours, function.  
	\end{minipage}	
\end{figure}

Second, I find no evidence of differences in experiences as leaders. To test this potential explanation, the survey asked current team leaders about their perspective on the size of the team and the frequency of conflict that team leaders at the firm face. The median leader reports a team size of 6-10 employees and reports experiencing conflict several times a year. I find no meaningful gender differences in responses by current team leaders that would suggest that men and women face different realities with respect to these features.  Figure \ref{fig:survey_perception}, Panel A 
shows that male and female leaders give similar answers with respect to team size (Bars 1 and 2) and the frequency of conflict (Bars 3 and 4). I also do not find that employees in lower-level positions perceive that female leaders are treated differently from male leaders.  When asked whether a hypothetical employee (whose first name was randomized to be either \textit{Matthias} or \textit{Sophie} to reflect a male or a female gender, respectively) is worried about whether they will have a good experience as a leader, especially with respect to conflict with team members or other leaders, I find no meaningful differences by either respondent gender or vignette gender (Panel B of Figure \ref{fig:survey_vignette}).

Third, while women in lower-level positions do not perceive female leaders to be treated differently, they appear to be less informed about what team leadership entails. When asked about the number of teammates for which team leaders at the firm are responsible, women are 29\% ($p$=0.000) more likely to think that teams are larger than 10 employees (Bars 1 and 2 of  Figure \ref{fig:survey_perception}, Panel A). This contrasts with the fact that the average team size at the firm is 6.  Women are also  12\% ($p$=0.000) more likely to think that negative interactions happen at least several times a month (Bars 3 and 4 of  Figure \ref{fig:survey_perception}, Panel B). This is more often than the frequency of conflict reported by current team leaders, which is several times a year. The gender gap in perceptions among employees in lower-level positions is in stark contrast to the absence of gender differences among current team leaders  (Figure \ref{fig:survey_perception}, Panel A). This finding suggests that there is a substantial share of women in lower-level positions who base their application decisions on overly negative perceptions of what team leadership entails. 

Taken together, my findings show that women who are at risk of first applying to leadership positions have overly negative perceptions of what team leadership entails and are more likely to be deterred by these features. In combination, these gender differences explain half of the gender application gap for team leadership positions (Column 5 of Appendix Table \ref{table:applied_pref}).  These findings highlight the importance of the design of team leadership positions for alleviating the gender leadership gap. Women's lower inclination to apply for promotions limits the already small internal applicant pools that firms rely on to fill higher-level positions (\citetalias{hkp2023}). In my sample, 52\% of higher-level positions receive zero internal female applicants, implying that the pool of female candidates firms can choose from if they were to attempt to increase female representation in leadership positions is limited.  Since leadership positions are characterized by large pay premiums---with senior leaders earning three times more than the average employee---the observed leadership gap also stands to impact realized earnings over employees' career. 

One implication of these results is that organizations seeking to increase female representation may be able to increase the appeal of these positions by modifying certain features of leadership positions. For example, offering leadership positions over smaller teams may increase interest among women. In addition, providing support to employees to help navigate the challenges related to leading a team may also encourage the pursuit of leadership positions. In the survey, women place particularly high values on mentoring and frequently express interest in learning how current leaders respond to conflict. Moreover, there appears to be scope for organizations to increase women's applications by assuaging overly negative perceptions of team leadership.  Strategies suggested in the survey include providing more information in job postings, such as information on the size and composition of the team, and access to job shadowing opportunities, which would offer an opportunity to directly observe the actual experience of team leaders.

\section{Conclusion}
\label{sec:conclusion}

This paper uses a new combination of personnel records, job application data, and large-scale employee surveys from a large multinational firm to test whether and why climbing the leadership ladder may be less appealing to women than to men.  I document large gender differences in applications for promotions for employees in lower-level positions. This application gap is not accounted for by  standard explanations, such as  risk preferences, confidence, and perceived application success.  Instead, I show that having to assume responsibility over a team, which is a salient feature of promotions, is less appealing to women than to men. The gender  gap with respect to team responsibility is large, robust, and translates into realized outcomes. 

My findings highlight the important role played by leadership design. While previous literature has documented critical gender differences with respect to job characteristics, such as flexibility or commutes, the requirement of taking responsibility over a team---which is a key prerequisite for career advancement in most organizations (\citetalias{hkp2023})---has not been studied. My results suggest that even in settings where women do not perceive differential treatment, large gender differences in applications for promotions arise because team leadership is less appealing to women than to men. Given the importance of responsibility over a team for career progression in organizations, the differential effect of leadership design has large consequences on  gender differences in representation and pay, and firms' ability to fill higher-level positions. 

Understanding the root causes of female underrepresentation is critical for identifying effective policy remedies. My findings highlight the importance of analyzing gender differences that arise early on in employees' careers when employees consider whether or not to sort into the leadership pipeline,  which have received less attention by previous research due to data limitations. By providing first evidence on the role of team leadership for gender differences in career outcomes, this paper motivates a series of important questions for future research on how leadership positions can be designed and advertised to be also appealing to women. For instance, my survey responses suggest that workers would value if firms provide additional information about the size and composition of the team in job postings. Workers also express a lot of interest in receiving advice on how to navigate the challenges of being a leader, for instance with respect to dealing with conflict or delegating administrative tasks.\footnote{There is particular interest (expressed by 37\% of workers) in learning from mentors who have a similar background and only recently got promoted.} 


\begin{singlespace}
	\bibliography{literature2023}
	\nocite{*}
\end{singlespace}

\clearpage
\newpage

\section{Figures and Tables}

\FloatBarrier
\FloatBarrier

\FloatBarrier
\FloatBarrier

\FloatBarrier
\FloatBarrier

\FloatBarrier
\FloatBarrier

\FloatBarrier
\FloatBarrier

\FloatBarrier
\FloatBarrier


\begin{table}[h]
	\caption{Summary Statistics of Quarterly Employee Sample}
	\begin{center}
		\scalebox{1.0}{{
\def\sym#1{\ifmmode^{#1}\else\(^{#1}\)\fi}
\begin{tabular}{l*{1}{ccc}}
\hline\hline
                    &         Men&       Women&    p-value        \\
                    & (1) & (2) & (3)\\
\hline
\textbf{Demographics}  &        &       &         \\
\hspace{0.2cm}German citizen          &       0.89&        0.84&        0.00\\
\hspace{0.2cm}Age (years)   &        43.9&        40.2&        0.00\\
\hspace{0.2cm}Education (years)    &   16.0&        14.9&        0.00\\
\hspace{0.2cm}Firm tenure (years)    &       13.3&        11.9&        0.00\\
\hspace{0.2cm}Married             &        0.64&        0.50&        0.00\\
\hspace{0.2cm}Children          &          0.77&        0.64&        0.00\\
\hspace{0.2cm}On parental leave   &        0.02&        0.08&        0.00\\
\textbf{Position}  &        &       &         \\
\hspace{0.2cm}Full-time        &         0.98&        0.74&        0.00\\
\hspace{0.2cm}Weekly hours      &        42.2&        37.9&        0.00\\
\hspace{0.2cm}Technical position & 0.72&        0.34&        0.00\\
\hspace{0.2cm}Salary decile       &          6.0&         3.5&        0.00\\
\hspace{0.2cm}Top management      &        0.08&        0.04&        0.00\\
\hspace{0.2cm}Team leadership            &        0.21&        0.09&        0.00\\
\hspace{0.4cm}Direct reports       &         6.1&         5.1&        0.00\\
\hspace{0.4cm}Cumulative reports  &         37.0&        10.7&        0.00\\
\hline
Observations        &     3xx,xxx &  1xx,xxx          &                \\
\hline\hline \multicolumn{4}{p{0.6\linewidth}}{\footnotesize Notes: This table reports descriptives statistics for the quarterly analysis sample from 2015 to 2019. Column 1 restricts to male employees, Column 2 restricts to female employees, and Column 3 presents the p-value from a test of equality of the two means. A technical position is defined as a job related to engineering, IT, quality management, or production. Top management positions include the C-suite, the corporate board, as well as high-level executive positions. The number of direct and cumulative reports is only calculated for employees with team leadership.}\\
\end{tabular}
}
}
	\end{center}
	\label{table:sample_descriptive}
\end{table}

\begin{table}[h]
	\caption{Gender Differences in Applications for Promotions: Different Measures}
	\begin{center}
		\scalebox{1.0}{{
\def\sym#1{\ifmmode^{#1}\else\(^{#1}\)\fi}
\begin{tabular}{l*{5}{c}}
\hline\hline
 \multicolumn{6}{l}{\textit{Outcome:} Application for Promotion}  \\
\hline
 \noalign{\vskip 2mm} 
   &\multicolumn{1}{c}{Combined}    &\multicolumn{1}{c}{Direct}&\multicolumn{1}{c}{Managerial}&\multicolumn{1}{c}{Reporting} &\multicolumn{1}{c}{Pay}\\
                      &\multicolumn{1}{c}{measure}    &\multicolumn{1}{c}{reports}&\multicolumn{1}{c}{autonomy}&\multicolumn{1}{c}{distance to CEO} &\multicolumn{1}{c}{increase}\\
                    &\multicolumn{1}{c}{(1)}&\multicolumn{1}{c}{(2)}&\multicolumn{1}{c}{(3)} &\multicolumn{1}{c}{(4)}&\multicolumn{1}{c}{(5)} \\
\hline
\noalign{\vskip 1mm} 
Female                 &      -0.297 &      -0.357&      -0.131&       0.041   &      -0.113   \\
  &     (0.071)            &     (0.061)         &     (0.035)         &     (0.048)        &     (0.055)            \\
\hline
Av ME for Women  &     -0.0015           &     -0.0024         &     -0.0028         &      0.0004          &     -0.0010          \\
Gender Gap in \%    &       -27.4         &       -30.2         &       -15.4         &         4.7       &        -14.7            \\
Observations      &      3xx,xxx         &        3xx,xxx       &     3xx,xxx    &        3xx,xxx      &        3xx,xxx                  \\

\hline\hline \multicolumn{6}{p{0.9\linewidth}}{\footnotesize \textit{Notes}:  This table documents gender differences in applications for promotions for employees in lower-level positions.  Each coefficient stems from a separate logit regression of an indicator for applying for a promotion on gender and a large set of controls. Each column uses a different measure to  identify promotions as higher-level positions with more job authority in the spirit of \citealp{Rosen1982} (see Section \ref{sec:hierarchy} for details).
	Column 1 defines promotions based on a combined measure of three dimensions of job responsibility---direct reports, managerial autonomy, and reporting distance to the CEO. Columns 2 to 4 separately use the three dimensions of job responsibility to define promotions as transitions to positions with more direct reports (Column 2),  with more managerial autonomy  over hours and business decisions (Column 3), and a closer reporting distance to the CEO (Column 4).  Column 5 defines a promotion as a transition to a position that yields a pay increase relative to employees' current position.  Controls: Age, German citizenship, educational qualification, marital status, family status, parental leave, firm tenure, function, division, location, full-time, hours, number of direct reports, and quarter fixed effects. Standard errors in parentheses.}\\ \end{tabular}}
}
	\end{center}
	\label{table:applied_promotions}
\end{table}

\begin{table}[h]
	\caption{Gender Differences in Applications for Promotions: Different Employee Groups}
	\begin{center}
		\scalebox{1.0}{{
\def\sym#1{\ifmmode^{#1}\else\(^{#1}\)\fi}
\begin{tabular}{l*{3}{c}}
\hline\hline
 \multicolumn{4}{l}{\textit{Outcome:} Application for Promotion}  \\
\hline
 \noalign{\vskip 1mm} 
      &\multicolumn{1}{c}{Lower }&\multicolumn{1}{c}{First Leadership }&\multicolumn{1}{c}{Any Leadership}\\

    &\multicolumn{1}{c}{Level }&\multicolumn{1}{c}{Level }&\multicolumn{1}{c}{Level}\\

                              &\multicolumn{1}{c}{(1) }&\multicolumn{1}{c}{(2) }&\multicolumn{1}{c}{(3)}\\
                    
\hline
Female              &      -0.297&       0.005         &      -0.036         \\
&     (0.071)         &     (0.170)         &     (0.169)         \\
\hline
Outcome Mean        &      0.0055         &      0.0048         &      0.0041         \\
Av ME for Women     &     -0.0015         &      0.0000         &     -0.0002         \\
Gender Gap in \%    &       -27.4         &         0.8         &        -5.2         \\
Observations   &          3xx,xxx           &       6x,xxx   &       7x,xxx          \\
\hline\hline \multicolumn{4}{p{0.7\linewidth}}{\footnotesize  \textit{Notes}: This table documents gender differences in applications for promotions for employees at different points of the leadership ladder. Column 1 focuses on employees in lower-level positions who work as individual contributors without enhanced job responsibility (hierarchy ranking of 20 or below). Column 2 focuses on employees in first-level leadership positions, for instance those who lead a team, head a project, or work as an expert with enhanced job responsibility (hierarchy ranking between 20 and 40). Column 3 focuses on employees in any type of leadership position, ranging from the first leadership level to top management positions up to the CEO (hierarchy ranking above 20).  Each coefficient stems from a separate logit regression of an indicator for applying for a promotion on gender and a large set of controls.  Promotions are defined using my preferred approach based on a combined measure of three dimensions of job authority (see Section \ref{sec:hierarchy}). Gender gaps in \% are computed by dividing the average marginal effect for women based on the logit coefficient by the outcome mean.  Controls: Age, German citizenship, educational qualification, marital status, family status, parental leave, firm tenure,  function, division,  location, full-time, hours, number of direct reports, and quarter fixed effects. Standard errors in parentheses.}\\ \end{tabular}}
}
	\end{center}
	\label{table:applied_main}
\end{table}

\begin{table}[h]
	\caption{Gender Differences in Applications for Different Job Transitions}
	\begin{center}
		\scalebox{1.0}{{
\def\sym#1{\ifmmode^{#1}\else\(^{#1}\)\fi}
\begin{tabular}{l*{8}{c}}
\hline\hline

\textit{Outcome}: &\multicolumn{4}{c}{ Lateral Switch}  &&\multicolumn{2}{c}{Promotion}    \\   \cline{2-5}  \cline{7-8} 
\noalign{\vskip 2mm} 
                              &\multicolumn{1}{c}{Any Lateral}&\multicolumn{1}{c}{Different}&\multicolumn{1}{c}{Different} &\multicolumn{1}{c}{Different}& &\multicolumn{1}{c}{Same}&\multicolumn{1}{c}{Different}   \\
                          &\multicolumn{1}{c}{Switch}&\multicolumn{1}{c}{Function}&\multicolumn{1}{c}{Division}&\multicolumn{1}{c}{City}  &&\multicolumn{1}{c}{City}&\multicolumn{1}{c}{City}  \\
                    &\multicolumn{1}{c}{(1)}&\multicolumn{1}{c}{(2)}&\multicolumn{1}{c}{(3)} &\multicolumn{1}{c}{(4)}    &&\multicolumn{1}{c}{(5)}&\multicolumn{1}{c}{(6)}\\
               
\hline
\noalign{\vskip 1mm} 
Female           &       0.132&       0.117         &       0.221&       0.139         &&      -0.316&      -0.246  \\
&     (0.057)         &     (0.085)         &     (0.081)         &     (0.106)         &&     (0.086)         &     (0.128)         \\
\hline
Outcome Mean       &      0.0065         &      0.0027         &      0.0029         &      0.0020         &&      0.0038         &      0.0018         \\
Av ME for Women     &      0.0009         &      0.0004         &      0.0008         &      0.0003         &&     -0.0011         &     -0.0004         \\
Gender Gap in \%    &        14.4         &        15.4         &        26.9         &        14.7         & &      -30.1         &       -21.3         \\
Observations   &      3xx,xxx         &       3xx,xxx          &       3xx,xxx        &       3xx,xxx          &&       3xx,xxx        &       3xx,xxx                 \\

\hline\hline \multicolumn{8}{p{0.9\linewidth}}{\footnotesize \textit{Notes}:  This table documents gender differences in applications for different types of transitions for employees in lower-level positions. Each column focuses on a different outcome variable. Column 1 focuses on applications for lateral switches that do not induce an increase in hierarchy ranking.  Column 2 focuses on applications for lateral switches to a different function relative to employees' current position (e.g. transition from engineering to sales). Column 3 focuses on applications for lateral switches to a different division at the firm relative to employees' current position.  Column 4 focuses on applications for lateral switches to a different city. 
Column 5 focuses on applications for promotions which would be in the same city as employee's current position. Column 6 focuses on applications for promotions which would be in a different city. Promotions are defined using my preferred approach based on a combined measure of three dimensions of job authority (see Section \ref{sec:hierarchy}). Each coefficient stems from a separate logit regression of an indicator for applying for a given transition on gender and a large set of controls. Gender gaps in \% are computed by dividing the average marginal effect for women based on the logit coefficient by the outcome mean.  Controls: Age, German citizenship, educational qualification, marital status, family status, parental leave, firm tenure, function, division,  location, full-time, hours, number of direct reports, and quarter fixed effects. Standard errors in parentheses.}\\ \end{tabular}}
}
	\end{center}
	\label{table:applied_lower_other}
\end{table}

\begin{table}[h]
	\caption{Heterogeneity in Gender Differences in Applications for Promotions}
	\begin{center}
		\scalebox{0.8}{{
\def\sym#1{\ifmmode^{#1}\else\(^{#1}\)\fi}
\begin{tabular}{l*{7}{c}}
\hline\hline
 \multicolumn{8}{l}{\textit{Outcome:} Application for Promotion}  \\
 \hline
 \noalign{\vskip 2mm} 
&\multicolumn{6}{c}{Employee Groups} \\ \cline{2-7} 
\noalign{\vskip 2mm} 
                          &\multicolumn{1}{c}{Works}&\multicolumn{1}{c}{No}&\multicolumn{1}{c}{Not}&\multicolumn{1}{c}{Female-Leaning} &\multicolumn{1}{c}{Female}&\multicolumn{1}{c}{High-Female}\\
                                           &\multicolumn{1}{c}{Full-Time}&\multicolumn{1}{c}{Children}&\multicolumn{1}{c}{Married}  &\multicolumn{1}{c}{Area}&\multicolumn{1}{c}{Supervisor}&\multicolumn{1}{c}{Team}\\
                    &\multicolumn{1}{c}{(1)}&\multicolumn{1}{c}{(2)}&\multicolumn{1}{c}{(3)} &\multicolumn{1}{c}{(4)}&\multicolumn{1}{c}{(5)}&\multicolumn{1}{c}{(6)} \\
\hline
\noalign{\vskip 1mm} 
Female               &      -0.322&      -0.216 &      -0.244 &      -0.704&      -0.519&      -0.345\\
&     (0.073)         &     (0.105)         &     (0.097)         &     (0.117)         &     (0.164)         &     (0.080)         \\
\hline
Outcome Mean        &      0.0058         &      0.0060         &      0.0057         &      0.0057         &      0.0069         &      0.0058         \\
Av ME for Women     &     -0.0020         &     -0.0013         &     -0.0013         &     -0.0037         &     -0.0031         &     -0.0018         \\
Gender Gap in \%    &       -34.0         &       -22.0         &       -23.6         &       -65.2         &       -44.8         &       -31.5         \\
Observations      &      3xx,xxx         &       9x,xxx         &      1xx,xxx    &       7x,xxx      &       2x,xxx        &      1xx,xxx           \\

\hline\hline \multicolumn{8}{p{1.0\linewidth}}{\footnotesize \textit{Notes}:  This table documents heterogeneity with respect to gender differences in applications for promotions for employees in lower-level positions. Each column focuses on a different employee group. Column 1 restricts to employees who work full-time. Column 2 restricts to employees who do not have children. Column 3 restricts to employees who are not married.  Column 4 restricts to employees who work in a female-leaning  area (i.e., HR, finance, marketing and sales, or purchasing). Column 4 restricts to employees who work under a female supervisor. Column 6 restricts to employees who work in a team with an above-median share of female teammates. Each coefficient stems from a separate logit regression of an indicator for applying for a promotion on gender and a large set of controls. Promotions are defined using my preferred approach based on a combined measure of three dimensions of job authority (see Section \ref{sec:hierarchy}). Gender gaps in \% are computed by dividing the average marginal effect for women based on the logit coefficient by the outcome mean.  Controls: Age, German citizenship, educational qualification, marital status, family status, parental leave, firm tenure, function, division, location, full-time, hours, number of direct reports, and quarter fixed effects. Standard errors in parentheses.}\\ \end{tabular}}
}
	\end{center}
	\label{table:applied_hetero}
\end{table}

\begin{table}[h]
	\caption{Heterogeneity in Gender Differences With Respect to Responsibility Over a Team}
	\begin{center}
		\scalebox{0.8}{{
\def\sym#1{\ifmmode^{#1}\else\(^{#1}\)\fi}
\begin{tabular}{l*{7}{c}}
\hline\hline
 \multicolumn{8}{l}{\textit{Outcome:} Wanting to Take on Responsibility over a Team}  \\
 \hline
 \noalign{\vskip 2mm} 
                  &\multicolumn{7}{c}{Employee Groups} \\ \cline{2-8} 
\noalign{\vskip 2mm} 
                     &\multicolumn{1}{c}{Works}&\multicolumn{1}{c}{No}&\multicolumn{1}{c}{Female-Leaning}&\multicolumn{1}{c}{High} &\multicolumn{1}{c}{High} &\multicolumn{1}{c}{High} &\multicolumn{1}{c}{Effective}\\
                  &\multicolumn{1}{c}{Full-Time}&\multicolumn{1}{c}{Children}&\multicolumn{1}{c}{Area}&\multicolumn{1}{c}{Risk} &\multicolumn{1}{c}{Competition} &\multicolumn{1}{c}{Confidence}&\multicolumn{1}{c}{Leader} \\
                      &\multicolumn{1}{c}{(1)}&\multicolumn{1}{c}{(2)}&\multicolumn{1}{c}{(3)}&\multicolumn{1}{c}{(4)}&\multicolumn{1}{c}{(5)}&\multicolumn{1}{c}{(6)}&\multicolumn{1}{c}{(7)}\\
\hline
Female                      &      -0.396&      -0.428&      -0.473&      -0.468&      -0.465&      -0.437&      -0.436\\
&    (0.0943)         &    (0.1167)         &    (0.1405)         &    (0.1305)         &    (0.1049)         &    (0.0985)         &    (0.0900)         \\
Outcome mean    &       0.249         &       0.228         &       0.254         &       0.281         &       0.287         &       0.241         &       0.286         \\
Female Coefficient &      -0.072         &      -0.075         &      -0.081         &      -0.089         &      -0.089         &      -0.073         &      -0.084         \\
Gender Gap in \%     &       -29.1         &       -32.8         &       -32.0         &       -31.8         &       -31.0         &       -30.4         &       -29.4         \\
Observations        &        4,xxx         &        2,xxx         &        1,xxx         &        2,xxx         &        3,xxx         &        4,xxx       &       4,xxx        \\
\hline\hline \multicolumn{8}{p{1.15\linewidth}}{\footnotesize \textit{Notes}:  This table documents heterogeneity in gender differences with respect to taking on responsibility over a team based on the employee survey. Each column focuses on a different group of employees in lower-level positions where the outcome of interest is an indicator for a respondent having chosen responsibility over a team as response to the question where they would like to see themselves with respect to their career in five years. Column 1 focuses on employees who work full-time. Column 2 focuses on employees without children. Column 3 focuses on employees who work in areas with a high female share (i.e., HR, finance, marketing and sales, and purchasing). Column 4 focuses on employees with high risk preferences ($\geq$ 7 out of 10) based on a question from \cite{dohmen2011}. Column 5 focuses on employees with a high willingness to compete ($\geq$ 7 out of 10) based on a question from  \cite{buser2021}. Column 6 focuses on employees with  high confidence ($\geq$ 5 out of 7) based on a question from \cite{rosenberg2015}. Column 7 focuses on employees who report feeling that they would be effective leaders. See Appendix Section \ref{sec:app_survey} for the exact question wording.  Each coefficient stems from a separate logit regression of an indicator for choosing responsibility over a team on gender and a set of controls. Gender gaps in \% are computed by dividing the average marginal effect for women based on the logit coefficient by the outcome mean.  Controls: Age, tenure, family status, hours, function.  Standard error in parentheses.}\\ \end{tabular}}
}
	\end{center}
	\label{table:survey_robustaspiration}
\end{table}

\clearpage
\newpage

\begin{table}[h]
	\caption{Gender Differences in Applications for Promotions at the Employee-Vacancy Level}
	\begin{center}
		\scalebox{1.0}{{
	\def\sym#1{\ifmmode^{#1}\else\(^{#1}\)\fi}
	\begin{tabular}{l*{3}{c}}
		\hline\hline

 \multicolumn{4}{l}{\textit{Outcome:} Application for Promotion}  \\
 \hline
 \noalign{\vskip 2mm} 
                    &\multicolumn{1}{c}{(1)}         &\multicolumn{1}{c}{(2)}         &\multicolumn{1}{c}{(3)}                             \\
                    	\noalign{\vskip 1mm} 
\hline
	\noalign{\vskip 1mm} 
Female        &      0.0042&      0.0071&      0.0071\\
                    &    (0.0019)         &    (0.0015)         &    (0.0015)         \\
Team Responsibility        &      0.0388&      0.0406&      0.0379\\
                    &    (0.0024)         &    (0.0023)         &    (0.0023)         \\
Female $\times$ Team Responsibility            &     -0.0361&     -0.0339&     -0.0339\\
                    &    (0.0036)         &    (0.0035)         &    (0.0035)         \\
\hline
	\noalign{\vskip 1mm} 
Employee Demographics                       &        X             &        X       &        X               \\
Current Position Characteristics                        &        X             &        X          &        X                    \\
Baseline Vacancy Characteristics          &          -        &        X         &        X           \\
Additional Vacancy Characteristics         &          -        &         -          &        X                  \\
\hline
Outcome Mean        &     0.028                               &              0.028                            &     0.028                                      \\
Observations                       &     39,xxx,xxx           &     39,xxx,xxx         &     39,xxx,xxx              \\
\hline\hline \multicolumn{4}{p{0.85\linewidth}}{\footnotesize \textit{Notes}: This table documents gender differences in applications for promotions at the employee-vacancy level for  employees in lower-level positions. Each column is based on a separate regression based on a linear probability model where the outcome of interest is an indicator for applying for a promotion in an employee’s choice set, scaled by 100. Promotions are defined using my preferred approach based on a combined measure of three dimensions of job authority (see Section \ref{sec:hierarchy}). Controls for employee demographics and current position characteristics include age, German citizenship, educational qualification, marital status, family status, parental leave, firm tenure, function, division, location, full-time, hours. Baseline vacancy controls capture a vacancy's function, division, location. Additional vacancy characteristics include required work experience, required degree,  required skills (problem solving, communication, analytical, intercultural, assertiveness, English proficiency), whether the position required frequent business travel, whether the job posting was available in English, and the length of the posting. Each regression also controls for  quarter fixed effects.  Robust standard errors are clustered at the employee level.}\\ \end{tabular}}
}
	\end{center}
\label{table:refpref_top}
\end{table}

\begin{table}[h]
	\caption{Testing for Alternative Factors at the Employee-Vacancy Level}
	\begin{center}
		\scalebox{0.8}{{
\def\sym#1{\ifmmode^{#1}\else\(^{#1}\)\fi}
\begin{tabular}{l*{5}{c}}
\hline\hline
 \multicolumn{6}{l}{\textit{Outcome:} Application for Promotion}  \\
 \hline
 \noalign{\vskip 2mm} 
                    &\multicolumn{1}{c}{(1)}         &\multicolumn{1}{c}{(2)}         &\multicolumn{1}{c}{(3)}     &\multicolumn{1}{c}{(4)}    &\multicolumn{1}{c}{(5)}             \\
\hline
Female              &      0.0047      &      0.0047 &     -0.0060&      0.0005         &      0.0050\\
                    &    (0.0019)                 &    (0.0019)         &    (0.0016)         &    (0.0019)         &    (0.0019)         \\
Team Responsibility$_j$ &      0.0379&    0.0388&      0.0399&      0.0398&      0.0394\\
                    &    (0.0024)                 &    (0.0024)         &    (0.0024)         &    (0.0024)         &    (0.0024)         \\
Female $\times$ Team Responsibility$_j$ &     -0.0351&     -0.0360&     -0.0317&     -0.0368&     -0.0364\\
                    &    (0.0036)                  &    (0.0036)         &    (0.0035)         &    (0.0037)         &    (0.0036)         \\
Z$_j$     &      0.0148    &    0.0004           &0.0265   &   0.0010   &           0.0047\\
                    &      (0.0030)            &     (0.0020)          &   (0.0027)     &    (0.0014)          &    (0.0012)           \\
     Female $\times$ Z$_j$  & -0.0158    &  -0.0074          &   0.0479  &   0.0339&      -0.0028        \\
                    &    (0.0049)        &    (0.0034)         &    (0.0055)    &    (0.0038)           &   (0.0024)              \\
\hline
\noalign{\vskip 1mm} 
 Z$_j$   &\multicolumn{1}{c}{A Lot Of}         &\multicolumn{1}{c}{$>$ BA}         &\multicolumn{1}{c}{Female-Leaning}     &\multicolumn{1}{c}{Female}    &\multicolumn{1}{c}{Many External}             \\

        &\multicolumn{1}{c}{Experience}         &\multicolumn{1}{c}{Degree}         &\multicolumn{1}{c}{Area}     &\multicolumn{1}{c}{Supervisor}    &\multicolumn{1}{c}{Applicants}             \\
\hline\hline \multicolumn{6}{p{1.1\linewidth}}{\footnotesize Notes: This table tests for alternative factors that may give rise to gender differences in applications and uses data at the employee-vacancy level for  employees in lower-level positions. Each column is based on a separate regression based on a linear probability model where the outcome of interest is an indicator for applying for a promotion in an employee’s choice set, scaled by 100. Promotions are defined using my preferred approach based on a combined measure of three dimensions of job authority (see Section \ref{sec:hierarchy}). To test for the importance of alternative factors, each regression includes an interaction of gender and the demeaned factor of interest $Z_j$. Column 1 controls for whether the job ad required many years of work experience. Column 2 controls for whether the job ad required more than a Bachelor's degree. Column 3 focuses on the gender composition of the vacancy's work environment and controls for whether the vacancy would be in a female-leaning area. Column 4 controls for the gender of the vacancy's supervisor. Column 5 controls for whether the vacancy received a high share of external applicants as measure of how appealing a vacancy appears, where high is defined by the top quartile among all vacancies in my sample. I control for quarter fixed effects, employee demographics and current position characteristics (age, German citizenship, educational qualification, marital status, family status, parental leave, firm tenure, function, division, location, full-time, hours) and baseline vacancy controls (function, division, location). Robust standard errors are clustered at the employee level. The number of observation across all regressions is 39,xxx,xxx. The outcome mean is 0.010.}\\
\end{tabular}
}
}
	\end{center}
	\label{table:refpref_alternatives}
\end{table}

\clearpage 
\newpage
\appendix
\appendixpage
\setcounter{table}{0}
\renewcommand{\thetable}{\thesection\arabic{table}}
\setcounter{figure}{0}
\renewcommand\thefigure{\thesection\arabic{figure}}

\section{Appendix Figures \& Tables}

\FloatBarrier
\begin{figure}[!h]
	\thisfloatpagestyle{plainlower}
	\caption{Example of Firm's Internal Job Portal}
	\centering
	\begin{minipage}[b]{0.8\linewidth}
		\centering
		\caption*{Panel A. Search Interface}
		\includegraphics[scale=0.2]{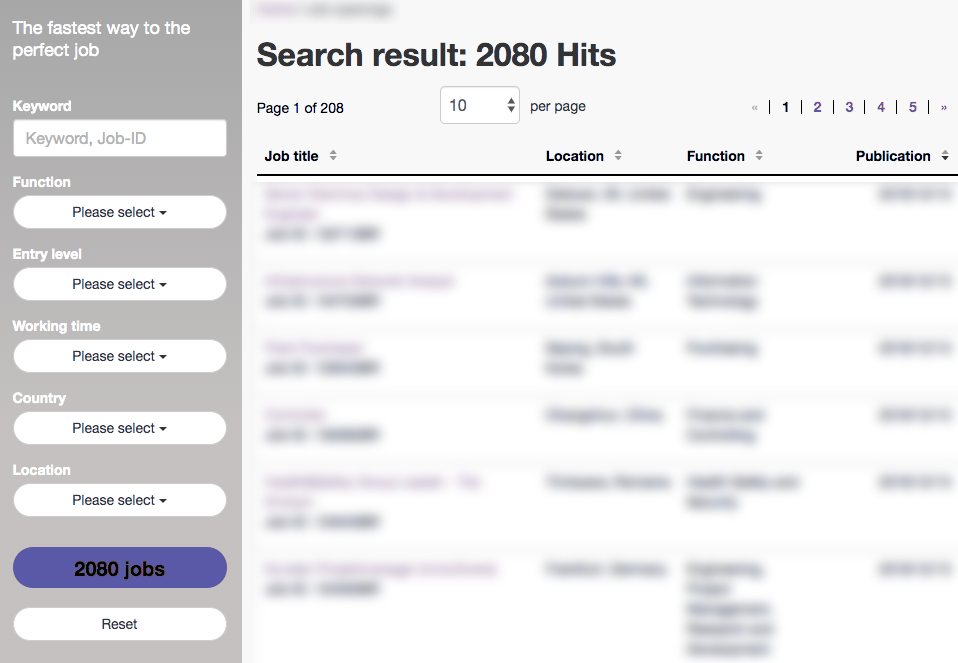}
	\end{minipage}	\\
	\vspace{1.5mm}
	\begin{minipage}[b]{0.8\linewidth}
		\centering
		\caption*{Panel B. Typical Job Ad}
		\includegraphics[scale=0.2]{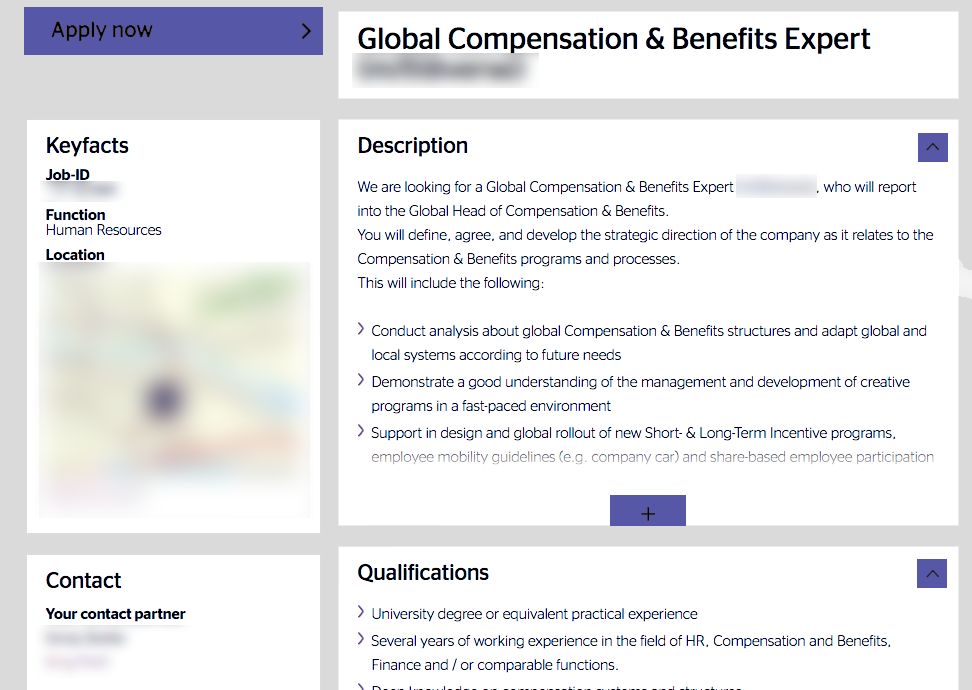}
	\end{minipage}	\\
	\vspace{1.5mm}
	\begin{minipage}[b]{0.8\linewidth}
		\centering
		\caption*{Panel C. Application Interface}
		\includegraphics[scale=0.22]{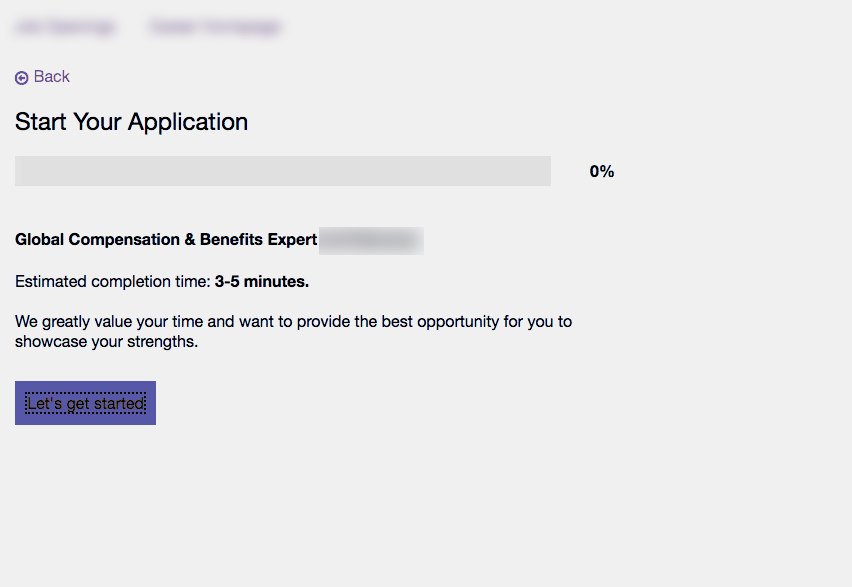}
	\end{minipage}		\\
	\label{fig:jobportal}
	\vspace{3mm}
	\begin{minipage}[b]{0.8\linewidth}
		\footnotesize \textit{Notes:} This figure provides an stylized example of the firm's internal job portal. Panel A displays the search interface,  Panel B illustrates a typical job ad, and Panel C presents the application interface through which employees submit  internal applications.  
	\end{minipage}	
\end{figure}
\FloatBarrier

\FloatBarrier
\begin{figure}[!ht]
	\thisfloatpagestyle{plainlower}
	\caption{Characteristics of the Combined Measure of Job Hierarchy}
	\vspace{0.5cm}
	\centering
	
	\begin{minipage}[b]{0.8\linewidth}
		\centering
		\caption*{Panel A: Distribution of Hierarchy Ranking}
		\includegraphics[scale=0.7]{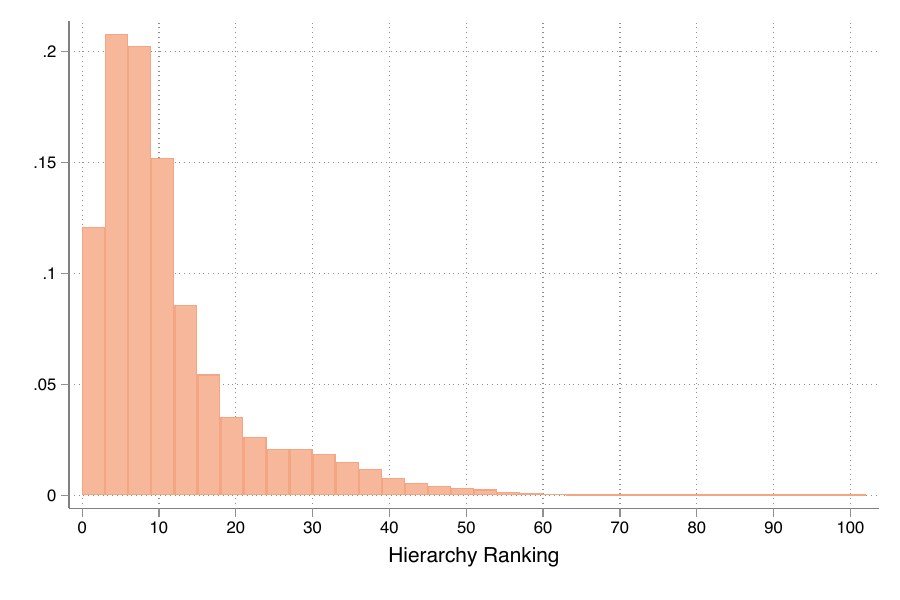}
	\end{minipage}	\\
	
	\vspace{0.5cm}
	\begin{minipage}[b]{0.8\linewidth}
		\centering
		\caption*{Panel B: Female Share by  Hierarchy Ranking}
		\includegraphics[scale=0.7]{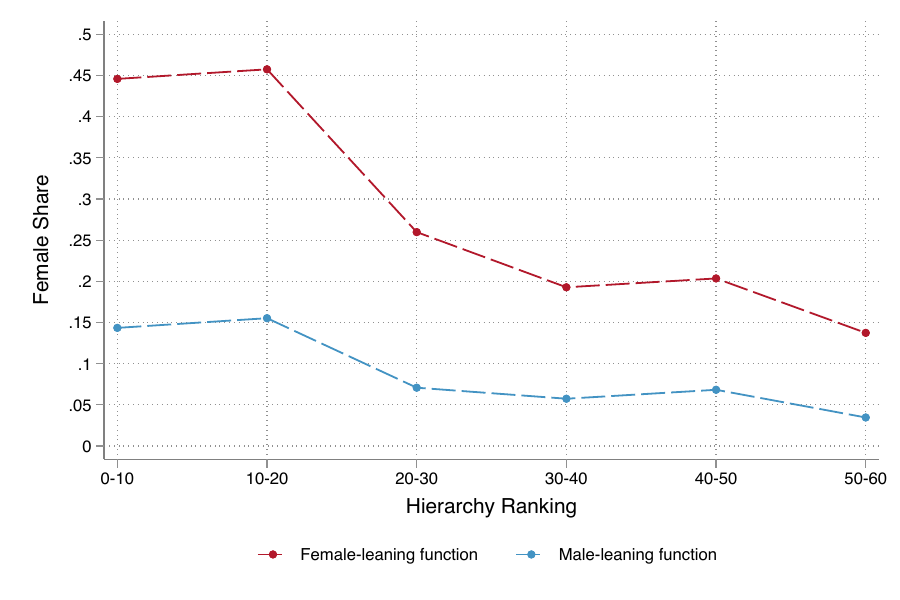}
	\end{minipage}	\\
	\label{fig:hierarchy_index_distribution}
	\begin{minipage}[b]{0.8\linewidth}
		\footnotesize \textit{Notes}: This figure provides descriptive statistics for measure of job hierarchy I construct. Panel A displays the distribution of the continuous hierarchy ranking in my sample. Low values represent lower-level positions, such as entry-level engineering jobs. The highest value of 100 represents the CEO. The majority of workers (84\%) are situated in positions with a hierarchy ranking of 20 or below. Panel B shows the female share by decile of the hierarchy ranking, separately for employees in female-leaning areas (i.e., human resources, finance, marketing and sales, and purchasing) and male-leaning areas (all remaining functions).  N= 4xx,xxx
	\end{minipage}	
\end{figure}
\FloatBarrier

\FloatBarrier
\begin{figure}[p]
	\thisfloatpagestyle{plainlower}
	\caption{Fit of Hierarchy Ranking Compared to Earnings}
	\centering
	\includegraphics[scale=0.6]{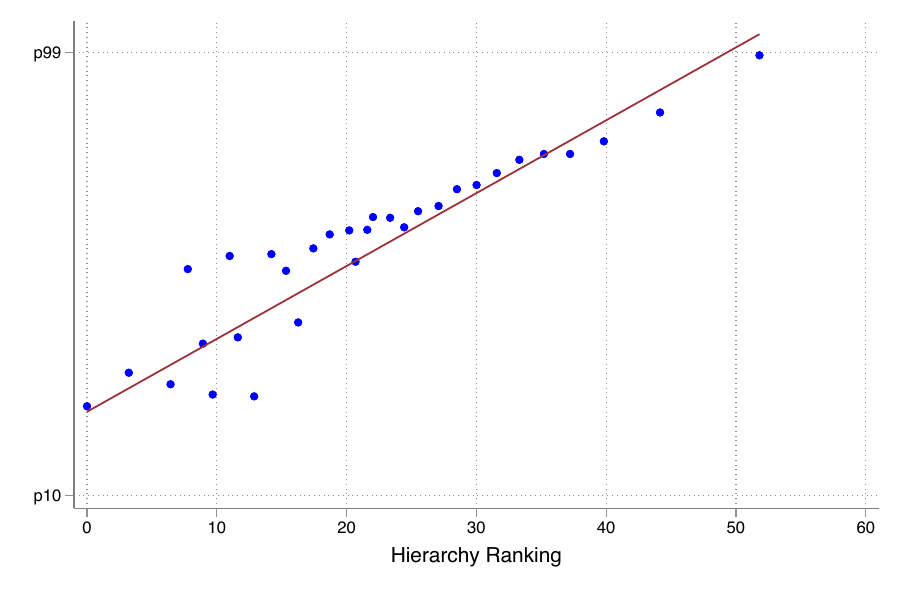}
	\label{fig:hierarchy_index_earningscorr}
	\begin{minipage}[b]{0.9\linewidth}
		\footnotesize \textit{Notes}: This figure displays  the correlation of the continuous hierarchy ranking and the percentile of workers' log real annual earnings at the firm.  To maintain confidentiality, earnings are displayed in terms of percentiles at the firm. I focus on hierarchy rankings below 60 because only a small minority of employees hold top management positions with rankings above 60 (e.g., CEO, board members).  N= 4xx,xxx 
	\end{minipage}	
\end{figure}

\begin{figure}[p]
	\thisfloatpagestyle{plainlower}
	\caption{Composition of Position Titles by Hierarchy Ranking}
	\centering
	\includegraphics[scale=0.7]{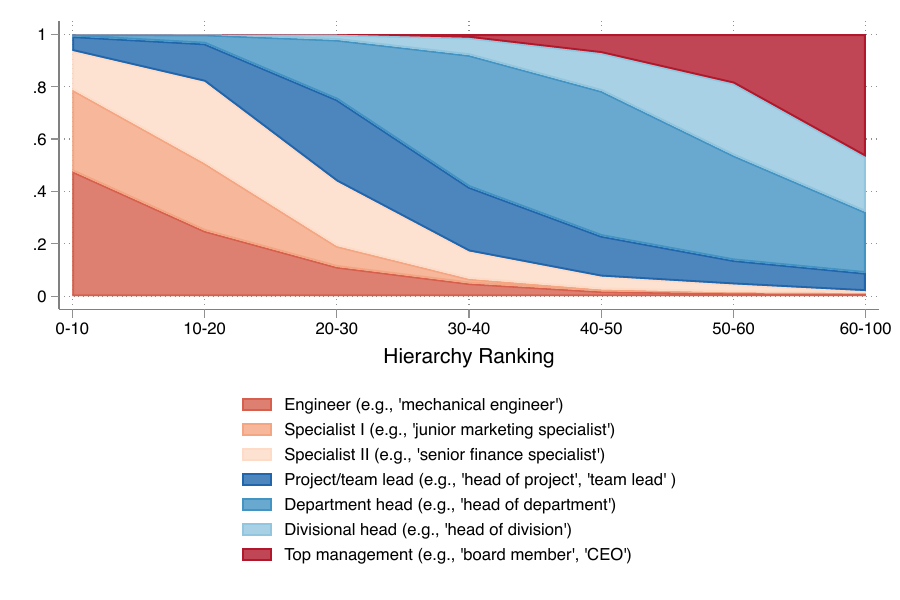}
	\label{fig:hierarchy_position_type}
	\begin{minipage}[b]{0.9\linewidth}
		\footnotesize \textit{Notes}: This figure displays the composition of position titles by deciles of the hierarchy ranking. I extract key terms from position titles that likely indicate the type of  responsibility that the position entails (e.g., ``engineer'', ``junior specialist'', ``head of department''). N=4xx,xxx
	\end{minipage}	
\end{figure}
\FloatBarrier

\FloatBarrier
\begin{figure}[!ht]
	\thisfloatpagestyle{plainlower}
	\caption{Employees' Job Search Activities}
	\centering
    \vspace{1mm} 
	\begin{minipage}[b]{0.8\linewidth}
		\centering
			\caption*{Panel A. Internal Job Search}
		\includegraphics[scale=3.4]{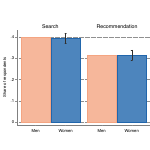}
	\end{minipage}	\\
\begin{minipage}[b]{0.8\linewidth}
	\centering
	\caption*{Panel B. External Job Search}
	\includegraphics[scale=3.4]{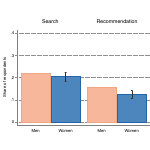}
\end{minipage}	\\
	\label{fig:survey2019_search}
	\begin{minipage}[b]{0.8\linewidth}
		\footnotesize \textit{Notes}:  This figure documents reported job search activities in the past 12 months based on the employee survey. Panel A shows the share of employees in lower-level positions who  searched for internal job openings and who received recommendations from others about internal job openings. Panel B shows the corresponding shares of searching for and receiving recommendations about external job openings. See Appendix Section \ref{sec:app_survey} for the exact question wording. For men, both panels show men's sample means. For women, outcome means plus the female coefficient are shown.  Controls: Age, tenure, family status, hours, function.  N=1x,xxx
	\end{minipage}	
\end{figure}
\FloatBarrier

\FloatBarrier
\begin{figure}[!ht]
	\thisfloatpagestyle{plainlower}
	\caption{Identity of Who Made Internal Recommendation}
	\centering
		\includegraphics[scale=0.7]{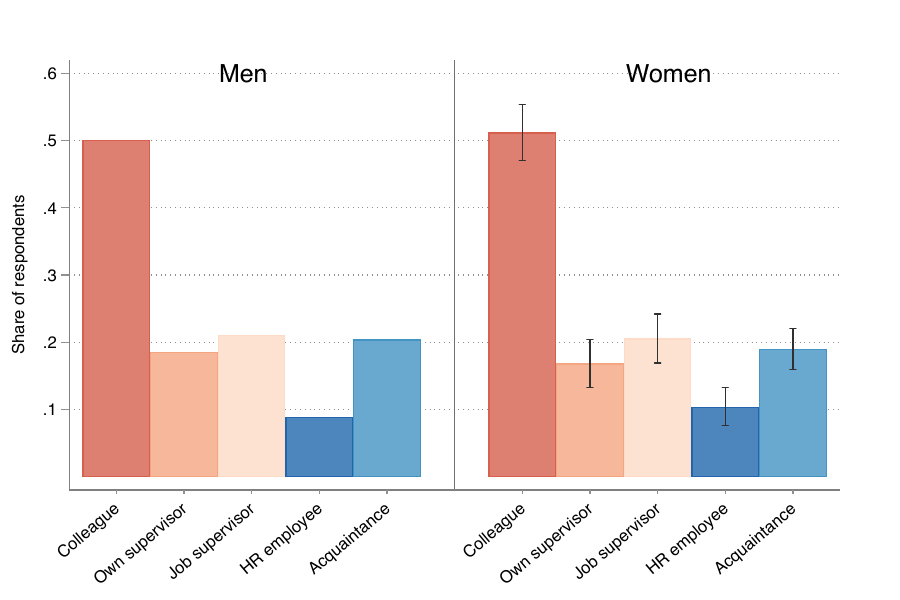}
	\label{fig:survey2019_rec}
	\begin{minipage}[b]{0.8\linewidth}
		\footnotesize \textit{Notes}:  This figure provides information about the job recommendations employees received in the past 12 months based on the employee survey. Employees who report having received a recommendation were asked to state the identity of the individual who made the recommendation. See Appendix Section \ref{sec:app_survey} for the exact question wording. For men, both panels show men's sample means. For women, outcome means plus the female coefficient are shown.  Controls: Age, tenure, family status, hours, function.  N=1x,xxx
	\end{minipage}	
\end{figure}
\FloatBarrier


\FloatBarrier
\begin{table}[!h]
	\caption{Comparison of Analysis Sample to Representative Survey of German Workforce}
	\begin{center}
		\scalebox{0.9}{{
\def\sym#1{\ifmmode^{#1}\else\(^{#1}\)\fi}
\begin{tabular}{l*{3}{c}}
\hline\hline
                   
                    &\multicolumn{1}{c}{\shortstack{Analysis Sample}}&\multicolumn{1}{c}{\shortstack{Large Manufacturer}}&\multicolumn{1}{c}{\shortstack{German Workforce}}\\
                     &\multicolumn{1}{c}{(1)}&\multicolumn{1}{c}{(2)}&\multicolumn{1}{c}{(3)}\\
\hline
Female                &        0.21&        0.21&        0.45\\
German citizen      &        0.88&        0.85&        0.88\\
Age (years)          &        43.1&        44.1&        43.9\\
Married           &        0.61&        0.59&        0.54\\
Children            &        0.80&        0.62&        0.62\\
Education (years)     &        15.7&        11.7&        11.8\\
Firm tenure (years) &        13.3&        16.8&        13.2\\
Weekly hours         &        41.1&        40.5&        38.1\\
Team leadership     &        0.18&        0.25&        0.29\\
\hline
Observations        &       9x,xxx&        1,472&       13,791\\
\hline\hline \multicolumn{4}{p{0.85\linewidth}}{\footnotesize \textit{Notes}: This table compares the employees in my sample to a representative survey of German workers (BIBB). Because the most recent version of the BIBB was conducted in 2018, I restrict my quarterly sample to 2018. Column 1 reports summary statistics for my analysis sample. Column 2 uses data from the BiBB survey and restricts to employees in manufacturing firms with at least 250 employees in West Germany. Column 3 uses data from the BiBB survey for all labor force participants in West Germany. Columns 2 and 3 are weighted to achieve representativeness of the German population. Team leadership is defined based on whether someone supervises direct reports. }\\ \end{tabular}}
}
	\end{center}
	\label{table:bibb_2018}
\end{table}

\begin{table}[!ht]
	\caption{Characteristics by Hierarchy Ranking}
	\begin{center}
		\scalebox{1.0}{{
\def\sym#1{\ifmmode^{#1}\else\(^{#1}\)\fi}
\begin{tabular}{c|cccc}
\hline\hline
                  Hierarchy  & Share&Tenure   & Share & Share   \\
                    Ranking & External Hire& (Years)&  $\geq$  BA Degree&   Bonus$/$Pay   \\
                    & (1) & (2) & (3) &(4)   \\
\hline
0-10                &       0.015&        11.6&        0.41&        0.12\\
10-20               &       0.011&        13.4&        0.54&        0.13\\
20-30               &       0.005&        16.2&        0.72&        0.16\\
30-40               &       0.002&        17.4&        0.78&        0.22\\
40-50               &       0.001&        17.3&        0.78&        0.30\\
50-60               &       0.000&        16.8&        0.79&        0.40\\
60-100              &       0.001&        16.1&        0.75&        0.40\\
\hline
N &\multicolumn{4}{c}{4xx,xxx} \\
\hline\hline
\multicolumn{5}{p{0.6\linewidth}}{\footnotesize \textit{Notes}: This table reports average characteristics by decile of the hierarchy ranking.  Column 1 reports the share of employees in my sample who join the firm from externally in a given quarter. Column 2 reports employees' tenure at the firm in years. Column 3 reports the share of employees who hold at least a Bachelor's degree.  Column 4 reports employees' bonus pay as fraction of total pay.}\\ 
\end{tabular}}}
	\end{center}
	\label{table:hierarchy_level_x}
\end{table}

\begin{table}
	\caption{Transition Matrix by Hierarchy Ranking}
	\begin{center}
		\scalebox{1.0}{{
\def\sym#1{\ifmmode^{#1}\else\(^{#1}\)\fi}
\begin{tabular}{c|ccccccc}
\hline\hline
&\multicolumn{6}{c}{Hierarchy Ranking$_{t+1}$ }\\
Ranking$_t$               &  0-10& 10-20 & 20-30&30-40&40-50&50-60\\
\hline
0-10&\textbf{68.5}&27.4&3.4&0.6&0.0&0.0\\
10-20&20.1&\textbf{58.6}&17.8&2.7&0.7&0.0\\
20-30&0.5&17.3&\textbf{50.0}&26.8&5.5&0.0\\
30-40&0.0&5.8&32.6&\textbf{45.3}&15.1&1.2\\
40-50&0.0&0.0&17.9&35.7&\textbf{35.7}&10.7\\
50-60&0.0&0.0&25.0&25.0&25.0&\textbf{25.0}\\
\hline\hline
\multicolumn{7}{p{0.6\linewidth}}{\footnotesize \textit{Notes}: This table represents the transition matrix for employees who switch positions between year $t$ and the subsequent year $t+1$. Each coefficient represents the share of employees (in \%) who start out at a hierarchy decile in year $t$ and transition to a respective decile in year $t+1$. For instance, 27.4\% of employees from the first hierarchy decile transition to a position in the subsequent decile. N$<$2,xxx }\\ \end{tabular}}}
	\end{center}
	\label{table:hierarchy_transition}
\end{table}

\begin{table}[h]
	\caption{Testing Robustness of the Application Gap: Definition of Lower-Level Positions}
	\begin{center}
		\scalebox{0.8}{{
\def\sym#1{\ifmmode^{#1}\else\(^{#1}\)\fi}
\begin{tabular}{l*{7}{c}}
\hline\hline
 \multicolumn{7}{l}{\textit{Outcome:} Application for Promotion}  \\
 \hline
      &\multicolumn{1}{c}{Direct}&\multicolumn{1}{c}{Managerial}&\multicolumn{1}{c}{Reporting}&\multicolumn{1}{c}{Pay} &\multicolumn{1}{c}{Combined measure}&\multicolumn{1}{c}{Combined measure}\\
&\multicolumn{1}{c}{reports}&\multicolumn{1}{c}{autonomy}&\multicolumn{1}{c}{distance to CEO}&\multicolumn{1}{c}{increase} &\multicolumn{1}{c}{cutoff 10}&\multicolumn{1}{c}{cutoff 30}\\
&\multicolumn{1}{c}{(1)}&\multicolumn{1}{c}{(2)}&\multicolumn{1}{c}{(3)}&\multicolumn{1}{c}{(4)}&\multicolumn{1}{c}{(5)}&\multicolumn{1}{c}{(6)}\\
\hline
\noalign{\vskip 1mm} 
Female              &      -0.354&      -0.136&       0.040         &      -0.133 &      -0.232&      -0.278\\
&     (0.058)         &     (0.035)         &     (0.068)         &     (0.091)         &     (0.080)         &     (0.066)         \\
\hline
Outcome Mean        &      0.0088         &      0.0205         &      0.0114         &      0.0068        &      0.0055         &      0.0057         \\
Av ME for Women     &     -0.0026         &     -0.0031         &      0.0006         &     -0.0010        &     -0.0011         &     -0.0015         \\
Gender Gap in \%    &       -30.0         &       -15.1         &         5.0         &       -14.4         &       -20.8         &       -26.6         \\
Observations      &      3xx,xxx         &        3xx,xxx          &      1xx,xxx      &        1xx,xxx       &      2xx,xxx          &      3xx,xxx            \\

\hline\hline \multicolumn{7}{p{1.2\linewidth}}{\footnotesize \textit{Notes}:  This table presents results from robustness exercises for employees in lower-level positions by using alternative approaches to identify these employees.  Each coefficient stems from a separate logit regression of an indicator for applying for a promotion on gender and a large set of controls. Each column uses a different measure to  identify promotions as higher-level positions with more job authority in the spirit of \citealp{Rosen1982} (see Section \ref{sec:hierarchy} for details).	Column 1 defines promotions as transitions to positions with more direct reports and uses an alternative definition of employees in lower-level positions as those without any direct reports. Column 2 defines promotions as transitions to positions with more managerial autonomy  over hours and business decisions and uses an alternative definition of employees in lower-level positions as those without any such autonomy, neither over hours nor business decisions. Column 3 defines promotions as transitions to positions with a closer reporting distance to the CEO and uses an alternative definition of employees in lower-level positions as those with the furthest distance to the CEO.  Column 4 defines promotions as  a transition to a position that yields a pay increase relative to employees' current position and uses an alternative definition of employees in lower-level positions as those whose current earnings are in the bottom quartile. Columns 5 and 6 use my preferred approach to define promotions based on the combined measure of three dimensions of job authority. Instead of defining lower-level positions as those with a hierarchy ranking below 20, Column 5 uses a cutoff of 10, while Column 6 uses a cutoff of 30. Controls: Age, German citizenship, educational qualification, marital status, family status, parental leave, firm tenure, function, division, location, full-time, hours, number of direct reports, and quarter fixed effects. Standard errors in parentheses.}\\ \end{tabular}}
}
	\end{center}
	\label{table:robust_group}
\end{table}

\begin{table}[h]
	\caption{Testing Robustness of the Application Gap: Controls and Sample Definitions}
	\begin{center}
		\scalebox{0.8}{{
\def\sym#1{\ifmmode^{#1}\else\(^{#1}\)\fi}
\begin{tabular}{l*{9}{c}}
\hline\hline
  \multicolumn{9}{l}{\textit{Outcome:} Application for Promotion}  \\
  \hline
  \noalign{\vskip 2mm}
  & \multicolumn{4}{c}{Controls} & & \multicolumn{3}{c}{Samples}  \\  \cline{2-5} \cline{7-9}
\noalign{\vskip 2mm} 
    &\multicolumn{1}{c}{Employee}&\multicolumn{1}{c}{Position}  &\multicolumn{1}{c}{Baseline +} &\multicolumn{1}{c}{Baseline} + &  &\multicolumn{1}{c}{Balanced}&\multicolumn{1}{c}{Applied}  &\multicolumn{1}{c}{Applied After} \\
        &\multicolumn{1}{c}{Demographics}&\multicolumn{1}{c}{Controls}  &\multicolumn{1}{c}{Performance} &\multicolumn{1}{c}{Time in Job} &  &\multicolumn{1}{c}{Sample}&\multicolumn{1}{c}{$\geq$ Once}  &\multicolumn{1}{c}{Rejection} \\
  &\multicolumn{1}{c}{(1)}&\multicolumn{1}{c}{(2)}&\multicolumn{1}{c}{(3)}   &\multicolumn{1}{c}{(4)} & &\multicolumn{1}{c}{(5)}&\multicolumn{1}{c}{(6)}   &\multicolumn{1}{c}{(7)}\\
  \hline
 Female              &      -0.351&      -0.389&      -0.214&      -0.298 &    &      -0.283&      -0.277&      -0.352\\
                    &     (0.062)         &     (0.066)         &     (0.073)         &     (0.071)     &   &     (0.082)         &     (0.076)         &     (0.120)     \\
\hline
Outcome Mean        &      0.0055         &      0.0055         &      0.0055         &      0.0055       &  &      0.0056         &      0.1928         &      0.0333   \\
Av ME for Women     &     -0.0018         &     -0.0021         &     -0.0013         &     -0.0015      &  &     -0.0014         &     -0.0345         &     -0.0087     \\
Gender Gap in \%    &       -33.3         &       -37.1         &       -22.8         &       -27.4  & &       -25.6         &       -17.9         &       -26.2        \\
  Observations   &      3xx,xxx           &      3xx,xxx           &      3xx,xxx         &      3xx,xxx     & &      2xx,xxx           &      1x,xxx           &      2x,xxx             \\

\hline\hline \multicolumn{9}{p{1.2\linewidth}}{\footnotesize \textit{Notes}:  This table presents results from robustness exercises for employees in lower-level positions. The outcome of interest are applications for promotions. Promotions are defined using my preferred approach based on a combined measure of three dimensions of job authority (see Section \ref{sec:hierarchy}).  Columns 1 to 4 use different sets of controls: only demographics (Column 1), only position information (Column 2), baseline controls plus performance ratings (Column 3), and baseline controls plus time in current job (Column 4). Columns 5 to 7 use baseline controls and distinguish between three subsamples of employees. Column 5 restricts to a balanced sample of employees who I observe in each quarter of my sample period. Column 6 restricts to those who applied at least once in a given quarter. Column 7 restricts those who applied after being rejected for an application in the past.  Each coefficient stems from a separate logit regression of an indicator for applying on gender. Gender gaps in \% are computed by dividing the average marginal effect for women based on the logit coefficient by the outcome mean. Baseline controls include age, German citizenship, educational qualification, marital status, family status, parental leave, firm tenure, function, division, location, full-time, hours, number of direct reports, and quarter fixed effects. Standard errors in parentheses.}\\
 \end{tabular}}
}
	\end{center}
	\label{table:robust_controls}
\end{table}

\begin{table}[h]
	\caption{Gender Differences in Applications for Promotions Using Pay: Different Employee Groups}
	\begin{center}
		\scalebox{0.8}{{
\def\sym#1{\ifmmode^{#1}\else\(^{#1}\)\fi}
\begin{tabular}{l*{3}{c}}
\hline\hline
 \multicolumn{4}{l}{\textit{Outcome:} Application for Promotion}  \\
 \hline
 \noalign{\vskip 1mm} 
      &\multicolumn{1}{c}{Lower }&\multicolumn{1}{c}{First Leadership }&\multicolumn{1}{c}{Any Leadership}\\

&\multicolumn{1}{c}{Level }&\multicolumn{1}{c}{Level }&\multicolumn{1}{c}{Level}\\

&\multicolumn{1}{c}{(1) }&\multicolumn{1}{c}{(2) }&\multicolumn{1}{c}{(3)}\\
\hline
Female           &      -0.113&       0.102         &       0.135         \\
&     (0.055)         &     (0.176)         &     (0.172)         \\
\hline
Av ME for Women       &     -0.0010         &      0.0008         &      0.0010         \\
Gender Gap in \%    &       -14.7         &        13.3         &        18.1         \\
Observations   &          3xx,xxx           &       6x,xxx   &       7x,xxx          \\
\hline\hline \multicolumn{4}{p{0.7\linewidth}}{\footnotesize  \textit{Notes}: This table documents gender differences in applications for promotions for employees at different points of the leadership ladder. Column 1 focuses on employees in lower-level positions who work as individual contributors without enhanced job responsibility (hierarchy ranking of 20 or below). Column 2 focuses on employees in first-level leadership positions, for instance those who lead a team, head a project, or work as an expert with enhanced job responsibility (hierarchy ranking between 20 and 40). Column 3 focuses on employees in any type of leadership position, ranging from the first leadership level to top management positions up to the CEO (hierarchy ranking above 20).  Each coefficient stems from a separate logit regression of an indicator for applying for a promotion on gender and a large set of controls.  Promotions are defined as a transition to a position that yields a pay increase  between 5\% and 30\% relative to employee's current position. Gender gaps in \% are computed by dividing the average marginal effect for women based on the logit coefficient by the outcome mean.  Controls: Age, German citizenship, educational qualification, marital status, family status, parental leave, firm tenure,  function, division,  location, full-time, hours, number of direct reports, and quarter fixed effects. Standard errors in parentheses.}\\ \end{tabular}}
}
	\end{center}
	\label{table:robust_pay}
\end{table}

\begin{table}[h]
	\caption{Success Likelihood Conditional on Applying for a Promotion}
	\begin{center}
		\scalebox{1.0}{{
\def\sym#1{\ifmmode^{#1}\else\(^{#1}\)\fi}
\begin{tabular}{l*{2}{c}}
\hline\hline
 \textit{Outcome:}   &\multicolumn{1}{c}{Interviewed}&\multicolumn{1}{c}{Hired}\\

 &\multicolumn{1}{c}{(1)}&\multicolumn{1}{c}{(2)}\\
\hline
Female              &       0.568&       0.525\\
&     (0.138)         &     (0.176)         \\
\hline
Outcome Mean        &      0.5479         &      0.1553         \\
Av ME for Women     &      0.1283         &      0.0700         \\
Gender Gap in \%    &        23.4         &        45.1         \\
Observations          &        2,xxx                 &        2,xxx               \\
\hline\hline \multicolumn{3}{p{0.6\linewidth}}{\footnotesize \textit{Notes}:  This table documents gender differences in success likelihood for employees in lower-level positions who applied for an internal promotion.  Promotions are defined using my preferred approach based on a combined measure of three dimensions of job authority (see Section \ref{sec:hierarchy}). Each coefficient stems from a separate logit regression of an indicator for getting invited for an interview or hired on gender and a large set of controls. Gender gaps in \% are computed by dividing the average marginal effect for women based on the logit coefficient by the outcome mean.  Controls: Age, German citizenship, educational qualification, marital status, family status, parental leave, firm tenure, function, division, location, full-time, hours,  number of direct reports, and quarter fixed effects. Standard errors in parentheses.}\\ \end{tabular}}
}
	\end{center}
	\label{table:hiring}
\end{table}

\begin{table}[h]
	\caption{Gender Differences in Entry to and Exit from Firm}
	\begin{center}
		\scalebox{1.0}{{
	\def\sym#1{\ifmmode^{#1}\else\(^{#1}\)\fi}
	\begin{tabular}{l*{5}{c}}
		\hline\hline
  \noalign{\vskip 2mm} 
		\textit{Outcome:}             &\multicolumn{2}{c}{Lower}  &   &\multicolumn{2}{c}{First Leadership} \\ 
     &\multicolumn{2}{c}{Level}  &   &\multicolumn{2}{c}{Level} \\\cline{2-3}  \cline{5-6}  
\noalign{\vskip 2mm}
&\multicolumn{1}{c}{Entry}&\multicolumn{1}{c}{Exit}&&\multicolumn{1}{c}{Entry}&\multicolumn{1}{c}{(Exit}\\
		&\multicolumn{1}{c}{(1)}&\multicolumn{1}{c}{(2)}&&\multicolumn{1}{c}{(3)}&\multicolumn{1}{c}{(4)}\\
		\hline
   Female  &      -0.108 &-0.176&  &       0.188         &          -0.239         \\
&     (0.044)               &     (0.060)         & &     (0.205)  &     (0.216)         \\
\hline
Outcome Mean        &      0.0138              &      0.0067         &&      0.0040    &      0.0051         \\
Av ME for Women     &     -0.0039        &     -0.0013         & &      0.0072         &     -0.0011         \\
Gender Gap in \%    &       -28.0         &       -19.8         &  &       177.9        &     -21.5         \\
Observations   &        3xx,xxx              &           3xx,xxx         &   &          6x,xxx    &         6x,xxx               \\
\hline\hline \multicolumn{6}{p{0.7\linewidth}}{\footnotesize \textit{Notes}:  This table documents gender differences in the likelihood of entry to and exit from the firm in a given quarter. Columns 1 and 2 focus on employees in lower-level positions who work as individual contributors without enhanced job responsibility. Columns 3 and 4 focus on employees in first-level leadership positions, for instance those who lead a team, head a project, or work as an expert with enhanced job responsibility. Each coefficient stems from a separate logit regression of an indicator for entry or exit on gender and a large set of controls. Gender gaps in \% are computed by dividing the average marginal effect for women based on the logit coefficient by the outcome mean. Controls: Age, German citizenship, educational qualification, marital status, family status, parental leave, firm tenure,  function, division,  location, full-time, hours, number of direct reports, and quarter fixed effects. Standard errors in parentheses.} \\ \end{tabular}}

}
	\end{center}
	\label{table:entryexit}
\end{table}

\begin{table}[h]
	\caption{Gender Differences in Reported Applications for Promotions}
	\begin{center}
		\scalebox{0.9}{{
	\def\sym#1{\ifmmode^{#1}\else\(^{#1}\)\fi}
	\begin{tabular}{l*{7}{c}}
		\hline\hline
  \multicolumn{7}{l}{\textit{Outcome:} Application for Promotion}  \\
  \hline
  \noalign{\vskip 2mm} 
		             & & \multicolumn{3}{c}{Employee Groups} & & \multicolumn{2}{c}{Additional Controls} \\ \cline{3-5} \cline{7-8}
\noalign{\vskip 2mm} 
				&\multicolumn{1}{c}{Full}&\multicolumn{1}{c}{High}&\multicolumn{1}{c}{High}&\multicolumn{1}{c}{High} &&\multicolumn{1}{c}{Admin Tasks +} &\multicolumn{1}{c}{+ Unpredicted}\\
				&\multicolumn{1}{c}{Sample}&\multicolumn{1}{c}{Risk}&\multicolumn{1}{c}{Competition}&\multicolumn{1}{c}{Confidence} &&\multicolumn{1}{c}{Potential Conflict} &\multicolumn{1}{c}{Overtime}\\

		&\multicolumn{1}{c}{(1)}&\multicolumn{1}{c}{(2)}&\multicolumn{1}{c}{(3)}&\multicolumn{1}{c}{(4)} &&\multicolumn{1}{c}{(5)}&\multicolumn{1}{c}{(6)}\\
		\hline
Female               &      -0.430&      -0.329 &      -0.378&      -0.387 &  &    -0.255  &      -0.232\\
&     (0.137)         &     (0.183)         &     (0.154)         &     (0.152)         & &    (0.138)         &     (0.138)         \\

\hline
Outcome Mean          &      0.0817         &      0.1255         &      0.1214         &      0.0941         &&      0.0817         &      0.0817         \\
Av ME for Women     &     -0.0277         &     -0.0341         &     -0.0354         &     -0.0288         & &    -0.0149         &     -0.0134         \\
Gender Gap in \%    &       -33.9         &       -27.2         &       -29.2         &       -30.6         &&       -18.2         &       -16.4         \\
		
Observations   &        6,xxx         &          2,xxx              &          3,xxx              &          4,xxx      & &        6,xxx  &        6,xxx                \\
\hline\hline \multicolumn{8}{p{1.1\linewidth}}{\footnotesize \textit{Notes}:  This table documents gender differences in applications for promotions for employees in lower-level positions based on the employee survey. Each coefficient stems from a separate logit regression of an indicator for applying for a promotion in form of a team leadership position on gender and a large set of controls. Gender gaps in \% are computed by dividing the average marginal effect for women based on the logit coefficient by the outcome mean.  Column 1  establishes the gender gap in applications for the full sample of survey respondents, controlling for employee demographics and position information. Column 2 focuses on employees with high risk preferences ($\geq$ 7 out of 10) based on a question from \cite{dohmen2011}. Column 3  focuses on employees with a high willingness to compete ($\geq$ 7 out of 10) based on a question from \cite{buser2021}. Column 4 focuses on employees with a high confidence ($\geq$ 5 out of 7) based on a question from \cite{rosenberg2015}. Columns 5 and 6 use the full sample and add additional controls based on information on how employees perceive uneappealing features of team leadership and to what extent they would trade-off these features against higher pay. See Appendix Section \ref{sec:app_survey} for the exact question wording and Section \ref{sec:consequences} for a detailed description of these features. Column 5 controls for the two most frequently named features: a high administrative burden and  a high potential of conflict. Column 6 adds the feature of unpredicted overtime as additional control.   Controls: Age, tenure, family status,  hours, function. Robust standard error in parentheses.}\\ \end{tabular}}}
	\end{center}
	\label{table:applied_pref}
\end{table}

\clearpage
\begin{table}[h]
	\caption{Unappealing Features of Team Leadership}
	\begin{center}
		\scalebox{1.0}{{
	\def\sym#1{\ifmmode^{#1}\else\(^{#1}\)\fi}
	\begin{tabularx}{\textwidth}{|l|X|}
		\hline\hline
   1. & ``Having to lead a team full of difficult and unhappy employees; I would like to continue to work on technical topics, not just conduct employee dialogues and take care of  admin tasks.'' \\

	2.	&``Having to spend all your time on HR tasks, which is a necessity if team sizes are very large.'' \\

   	3.	& ``Purely administrative tasks.'' \\

   	4.	& ``Bureaucracy and complex and time-consuming approval processes.'' \\

	5.	& ``Admin and bureaucracy which leaves little time for operative tasks.'' \\

6.		&``With large teams, it’s impossible to do justice to your responsibility to lead the team while also having enough time to complete the actual tasks required of you.'' \\

	7.			&``Leadership position for a team of  10 or more: many meetings and very little  time for deep work.'' \\

    8.	& ``Constantly chasing people to do their tasks.'' \\

	9.	&``Leading a large team which doesn’t leave any time for good people management.'' \\

 10.	& ``Leading a large team of lone wolves.'' \\

	11.	& ``Having to rate employees instead of developing them; the task of laying off employees. '' \\

12. &``Situations in which I have to convey unpleasant decisions to the team: Disciplinary measures, reprimands, layoffs, refusal of vacation requests.''\\

13. &``Conversations with difficult employees, team leadership positions are often lonely.'' \\

14. &``You are often the first point of contact for your employees' problems, both professional and personal.'' \\

15. & ``Having to handle and withstand conflict with team members.'' \\
16. & ``Your main job is to deal with negative situations.'' \\

17.& ``Being caught between upper management and your team and having to constantly disappoint your own people.''\\

18.& ``Having to deal with increased potential for conflict between various stake holders.'' \\

19.	&``These are jobs that offer no room for personal development, but their only task is to push uncomfortable instructions from the top to the bottom.'' \\

20.	&``High stress: A lot of pressure by the team and a lot of pressure by the own supervisor.'' \\

\hline\hline \multicolumn{2}{p{0.95\linewidth}}{\footnotesize \textit{Notes}: This table displays quotations from the employee survey that describe employees' perception of what responsibility over a team entails at the firm. The quotations are based on free-text responses to the question which dimensions of responsibility over a team are least appealing. See Appendix Section \ref{sec:app_survey} for the exact question wording.}\\ \end{tabularx}}

}
	\end{center}
	\label{table:survey_quotes}
\end{table}


\clearpage
\newpage
\setcounter{table}{0}
\renewcommand{\thetable}{\thesection\arabic{table}}
\setcounter{figure}{0}
\renewcommand\thefigure{\thesection\arabic{figure}}

\section{Data Appendix}
\label{sec:appendix_data}
This section provides additional information on the data used in this paper. To comply with the firm's data protection regulations, the main dataset was constructed  based on the firm's personnel records and  internal application data, and anonymized in a secure data environment at the firm. In addition, the paper also  uses data from large-scale surveys I designed and fielded to the firm's employees.

\subsection{Personnel Records}

Spell-level information from the firm's personnel records  was collected that provides information on entry to and exit from the firm, employment status (e.g., regular employees), age, citizenship, educational qualifications, martial status, family status, parental leave status,  occupational code, function, division, location, full-time status, weekly hours, id of direct supervisor, managerial autonomy, base pay, bonus pay, performance ratings, potential ratings.  The final dataset restricts to regular employees only. The personnel records were collapsed to a quarterly employee dataset, keeping the the first observation of each employee in a given quarter. Based on the identity of the supervisor, several additional variables were created, for instance the number of direct report or the female share of the team. 

\subsection{Job Vacancy and Application Data}
From the firm's internal application platform, information on each job vacancy and each  applicant was collected. The collected information includes the time a vacancy was posted, the timing of each application, the outcome of an application (e.g., rejection, interview, offer), the id of the supervisor of the vacancy, and a wide range of vacancy characteristics that include for instance function, division, location, full-time status, required work experience, required degree,  required skills (problem solving, communication, analytical, intercultural, assertiveness, English proficiency), whether the position required frequent business travel, whether the job posting was available in English, the length of the posting (number of characters), the extent of team leadership, and managerial autonomy. Based on the identity of the vacancy's supervisor and the timing of the vacancy, several additional characteristics were constructed, for instance the gender of the supervisor. For each vacancy, the number of all external applicants (who were not employed at the firm at the time of application) was collected. The application histories of internal applicants were collapsed to a quarterly level and matched to the quarterly personnel records.

\subsection{Employee Survey}
\label{sec:app_survey}
Employees in my analysis sample were invited via e-mail by the firm's human resources department. Following the common practices of employee surveys at the firm, respondents were not incentivized to participate in the survey. However, the invitation highlighted that by participating in the survey, employees had the opportunity to provide their feedback on the internal labor market at the firm. The extent of positive feedback that employees provided afterwards about the survey confirms employees' positive view of the survey. Both survey waves  in Germany received a 50\% response rate, suggesting that survey participation was indeed appealing to employees.\footnote{To probe the robustness of my results, I also conducted a survey with white-collar and management employees of the firm that are based in the United States. This survey received a 36\% response rate.} 

While survey responses cannot be linked to administrative employee records at the individual level due to stringent data protection regulations in Germany, I assess the accuracy of the survey responses by comparing aggregated patterns from the survey to those observed in administrative data.  I find very high similarities both in worker demographics and position characteristics (Appendix Table \ref{table:survey_X}). In addition, I do not find any evidence for differential response propensities by gender when examining whether employees respond before a reminder was sent out (Appendix Table \ref{table:survey_reminder}). In unreported results, I find that employees who responded before the reminder was sent have similar observable characteristics compared to those who responded afterwards. I also find similar gender differences in my main outcomes for both sets of respondents. In addition, the patterns observed in the survey data highly resemble realized outcomes in the administrative data. For instance, while the administrative data suggest that women in lower-level positions are 27.4\% less likely to apply for early-career promotions, the corresponding difference is  33.9\% in the survey. I also find that patterns across both surveys in Germany, which were advertised using different wording, are highly similar.

\FloatBarrier

\begin{table}
	\caption{Comparison of Analysis Sample to Respondents of Employee Surveys}
	\begin{center}
				\scalebox{0.8}{{
\def\sym#1{\ifmmode^{#1}\else\(^{#1}\)\fi}
\begin{tabular}{l*{3}{c}}
\hline\hline
                    &\multicolumn{1}{c}{Sample}&\multicolumn{1}{c}{Survey 1}&\multicolumn{1}{c}{Survey 2}\\
                                        &\multicolumn{1}{c}{(1)}&\multicolumn{1}{c}{(2)}&\multicolumn{1}{c}{(3)}\\
\hline
Age (years)& & & \\
\hspace{0.2cm}$<$30          &        0.10&        0.11&        0.07\\
\hspace{0.2cm}30-39                &        0.31&        0.27&        0.25\\
\hspace{0.2cm}40-49        &        0.30&        0.22&        0.24\\
\hspace{0.2cm}$\geq$50           &        0.29&        0.24&        0.26\\
Firm tenure (years)& & & \\
\hspace{0.2cm}$<$3          &        0.10&        0.12&        0.04\\
\hspace{0.2cm}3-5          &        0.12&        0.15&        0.12\\
\hspace{0.2cm}6-9            &        0.13&        0.15&        0.15\\
\hspace{0.2cm}$\geq$10       &        0.55&        0.44&        0.51\\
Children     &        0.74&        0.58&        0.63\\
Weekly hours        &       41.30&       40.47&       40.39\\
Function& & & \\
\hspace{0.2cm}Engineering          &        0.49&        0.36&        0.31\\
\hspace{0.2cm}Finance              &        0.08&        0.05&        0.05\\
\hspace{0.2cm}HR            &        0.03&        0.03&        0.04\\
\hspace{0.2cm}IT               &        0.05&        0.04&        0.05\\
\hspace{0.2cm}Marketing and Sales &        0.10&        0.06&        0.06\\
\hspace{0.2cm}Logistics          &        0.05&        0.04&        0.04\\
\hspace{0.2cm}Purchasing            &        0.04&        0.03&        0.03\\
\hspace{0.2cm}Quality               &        0.07&        0.07&        0.06\\
\hline
Observations        &       4xx,xxx&       1x,xxx&        1x,xxx\\
\hline\hline \multicolumn{4}{p{0.6\linewidth}}{\footnotesize \textit{Notes}: This table compares average characteristics of my main analysis sample (Column 1) to the subset of employees who responded to the first employee survey (Column 2) and those who responded to the second employee survey (Column 3). }\\
\end{tabular}
}
}
	\end{center}
	\label{table:survey_X}
\end{table}

\begin{table}
	\caption{Selection into Survey Response by Gender}
	\begin{center}
		\scalebox{1.0}{{
\def\sym#1{\ifmmode^{#1}\else\(^{#1}\)\fi}
\begin{tabular}{l*{2}{c}}
\hline\hline
 &\multicolumn{1}{c}{Survey 1}         &\multicolumn{1}{c}{Survey 2}         \\
                    &\multicolumn{1}{c}{(1)}         &\multicolumn{1}{c}{(2)}         \\
\hline
Female              &       0.057         &      -0.034         \\
&    (0.0481)         &    (0.0550)         \\
\hline
Outcome mean        &       0.616         &       0.550         \\
Av ME for Women     &       0.013         &      -0.008         \\
Gender Gap in \%    &       2.0      &        -1.5         \\
Observations        &       1x,xxx         &         1x,xxx           \\
  \hline\hline \multicolumn{3}{p{0.6\linewidth}}{\footnotesize Notes: This table tests differential selection into survey response by gender. Estimates stem from a logit regression of completing the survey before reminders were sent out on gender and employee controls.  Controls: Age, tenure, family status, hours, function. Standard errors in parentheses.}\\ \end{tabular}}}
	\end{center}
	\label{table:survey_reminder}
\end{table}

\FloatBarrier

\noindent The translation of the relevant questions for this study is presented in abbreviated format:
\vspace{0.3cm}

\noindent \textbf{A.} In the past 12 months, have you applied for a position that requires responsibility over a team at   \{FIRM\}? 
  \{Yes, No\}

\vspace{0.3cm}
\noindent \textbf{B.} What best describes where would you like to see yourself in five years with respect to your career?
I would like to see myself in a position that …
  \{is similar to my current position, pays more, 	has more responsibility over a team, has more responsibility over a project, offers more challenges, has more flexible hours\}  

\vspace{0.3cm}
\noindent \textbf{C.} Are you generally a person who is willing to take risks or do you try to avoid taking risks?  Please choose a value on the scale below, where the value 0 means `not at all willing to take risks’ and the value 10 means `very willing to take risks’.
  \{0 (not at all willing to take risks), 1, 2, 3, 4, 5, 6, 7, 8, 9, 10 (very willing to take risks)\}  
  
  \vspace{0.3cm}
\noindent \textbf{D.} How competitive do you consider yourself to be? 
Please think about situations in which you would be in competition with others (e.g., applications).  Please choose a value on the scale below, where the value 0 means `not competitive at all’ and the value 10 means `very competitive’.
  \{0 (not competitive at all), 1, 2, 3, 4, 5, 6, 7, 8, 9, 10 (very competitive)\}

  \vspace{0.3cm}
\noindent \textbf{E.} Next, we would like to know which type of position is most appealing to you. If you were to apply for a new position in the next 12 months, which position would you prefer?

\noindent \{Position A: Similar to your current position + offers leadership of a small team,  \\
Position B: Similar to your current position + does not involve team leadership\} 

  \vspace{0.3cm}
\noindent \textbf{F.} Among the following options, which position would you prefer

\noindent \{Position A: Pays 10\% more than your current position + involves leading a small team,  \\
Position B: Pays 30\% more than your current position + involves leading a large team\} 

  \vspace{0.3cm}
\noindent \textbf{G.} Which position would you prefer now?

\noindent \{Position A: Pays 30\% more than your current position + often requires overtime,  \\
Position B: Pays 10\% more than your current position + has very predictable hours\}

  \vspace{0.3cm}
\noindent \textbf{H.} Finally, which position would you prefer?

\noindent \{Position A: Pays 10\% more than your current position + involves leading a team that is known to be very collegial,  \\
Position B: Pays 30\% more than your current position + involves leading a team that you do not know\} 

  \vspace{0.3cm}
\noindent \textbf{I.} At  \{FIRM\}, different programs for how to support employees in their future career planning are currently under consideration. To design the best programs, we would like to know what types of support employees like you find helpful.  If you could participate in one career workshop at your  \{FIRM\}  location, which workshop would you be most interested in?
  \{“How to create a career without responsibility over a team”, “How to become a successful team leader”,  “How to navigate work-life balance”, I am not interested in participating in any of these workshops\}

  \vspace{0.3cm}
\noindent \textbf{J.}** When thinking back to your own transition from individual contributor to a higher-level position, what were the most salient changes that accompanied this transition? Please select all that apply. 
  \{Started reporting to employees at higher levels, More responsibility over a team, More responsibility over projects, More responsibility over critical business decisions, Increase in regular working hours or business travel, Increase in unpredictable overtime, Change in working environment (e.g., different co-workers, work culture)\}

 \vspace{0.3cm}
\noindent \textbf{K.} Which features or typical tasks of team leadership at \{FIRM\} do you think you would find least appealing? \\
\{Free-form text response\}

 \vspace{0.3cm}
\noindent \textbf{L.} What do you think, how many employees typically report to a team leader at \{FIRM\}?
 \{1-2, 3-5, 6-10, 11-20, 21-49, 50+\}  

 \vspace{0.3cm}
\noindent \textbf{M.} In a typical week, how often do you think team leaders experience an unpredictable event that causes them to work longer or different hours than usual?
 \{Almost never, 1-2 days a week, 3 or more days a week\}  

 \vspace{0.3cm}
\noindent \textbf{N.} How often do team leaders encounter resistance from employees (e.g., because they had to make an unpopular decision)?
 \{A few times a year, Several times a year, Several times a month, Several times a week\}   

 \vspace{0.3cm}
\noindent \textbf{O.} We are now interested in your own personal experience. To what extent do you agree with the following statement? “I have confidence in my capabilities.” Please choose a value on the scale below, where the value 1 means ‘totally disagree’ and the value 7 means ‘totally agree’.
 \{1 (totally disagree), 2, 3, 4, 5, 6, 7 (totally agree)\}  

\vspace{0.3cm}
\noindent \textbf{P.} Next, please imagine how it would be if you were to hold a leadership position at \{FIRM\}. How much do you think you would enjoy leading a team?
 \{Not at all, Not, A bit, Much, Very much\}  

 \vspace{0.3cm}
\noindent \textbf{Q.}** How much do you enjoy leading a team?
 \{Not at all, Not, A bit, Much, Very much\}  

 \vspace{0.3cm}
\noindent \textbf{R.}  Compared to others, how would you rate your ability to be an effective leader?
 \{Very low, Low, High, Very high\}

 \vspace{0.3cm}
\noindent \textbf{S.}  Based on your own experiences or experiences others shared with you, we are now interested in your thoughts on helpful career advice for employees at \{FIRM\}. Imagine getting approached by S\{Sophie/Matthias\} who has been working in \{her/his\} position as specialist at \{FIRM\} for two years and has received high performance ratings and is considering applying for a team leader position.

\noindent If an employee like \{Sophie/Matthias\} were to apply for a team leadership position through \{FIRM\}’s online job portal, how large do you think is the probability they would get hired?
 \{0\% (definitely not), 10\%, 20\%, 30\%, 40\%, 50\%, 60\%, 70\%, 80\%, 90\%, 100\% (definitely yes)\}  

 \vspace{0.3cm}
\noindent \textbf{U.}  \{Sophie/Matthias\} is worried about whether \{she/he\} will have a good experience as a leader at \{FIRM\}, especially with respect to encountering conflict with team members or other leaders. What best describes how you would address \{her/his\} worry?

\noindent I think that employees like \{Sophie/Matthias\} who for the first time take on team leadership typically … 
 \{Do not enjoy being a leader, Have a difficult start, but eventually find their footing, Have many more good than bad moments, Really enjoy being a leader from the beginning\}  

 \vspace{0.3cm}
\noindent \textbf{T.} In the past, what were the main reasons you did not apply for a team leadership position at \{FIRM\}? \{Free-form text response\}

 \vspace{0.3cm}

\noindent \textbf{V.} Have you actively searched for an open position (internal or external) in the last 12
months?  \{Yes, No\}  

 \vspace{0.3cm}
 
\noindent \textbf{W.} Did someone recommend an open position (internal or external) to you in the past 12
months? \{Yes, No\}  

 \vspace{0.3cm}
 
\noindent \textbf{X.} You indicated that someone recommended one or more open positions to you. Who
recommended these positions to you? Please select all that apply. \{Colleagues within your team, Colleagues outside of your team, Your supervisor, The supervisor responsible for the open position, HR, Family or friends outside of \{FIRM\}, Other\}  \\

 \vspace{0.3cm}
 
\noindent** Only asked of employees who currently hold a leadership position

\clearpage 
\newpage

\section{Complementary Results}
This section provides three sets of complementary results. First, I provide an analogous analysis of potential gender differences for employees who currently already hold a leadership position. Second, I present additional results of constant availability as a potential salient feature of how team leadership positions are designed. Third, I probe the external validity of my results by discussing results from an analogous survey of the firm's employees based in the United States. 

\subsection{Gender Differences Among Leaders}
\label{sec:appendix_leaders}
Motivated by the large gender gap in applications for promotions among employees in lower-level positions, the majority of this paper focuses on employees in these lower-level positions.  While interpreting gender differences across different stages of the leadership ladder is difficult due to the possibility of differential self-selection, whether or not such gender differences exist at higher levels of the leadership ladder has key policy implications.


Building on Table \ref{table:applied_main} that shows the absence of meaningful differences in applications among employees who already hold leadership positions, this section examines additional dimensions along which critical gender differences among employees in leadership positions may occur. Drawing on the administrative records, I find that in addition to not being less likely to apply for promotions, women are also not less likely to be interviewed or hired conditional on applying (Appendix Table \ref{table:hiring_top}). I also find no evidence that women in leadership positions report lower career aspirations relative to men. Panel A of Appendix Figure \ref{fig:survey_leaders} shows that men and women in leadership positions are similarly likely to aspire holding more responsibility over a team in the next five years. This result differs substantially from the large gender gap with respect to team leadership that I document for employees in lower-level positions.

\begin{figure}[!ht]
	\thisfloatpagestyle{plainlower}
	\caption{Leaders' Perceptions of Differential Treatment}
	\centering
		\vspace{0.2cm}
	\begin{minipage}[b]{0.8\linewidth}
		\centering
			\caption*{Panel A. High Hiring Likelihood}
		\includegraphics[scale=0.8]{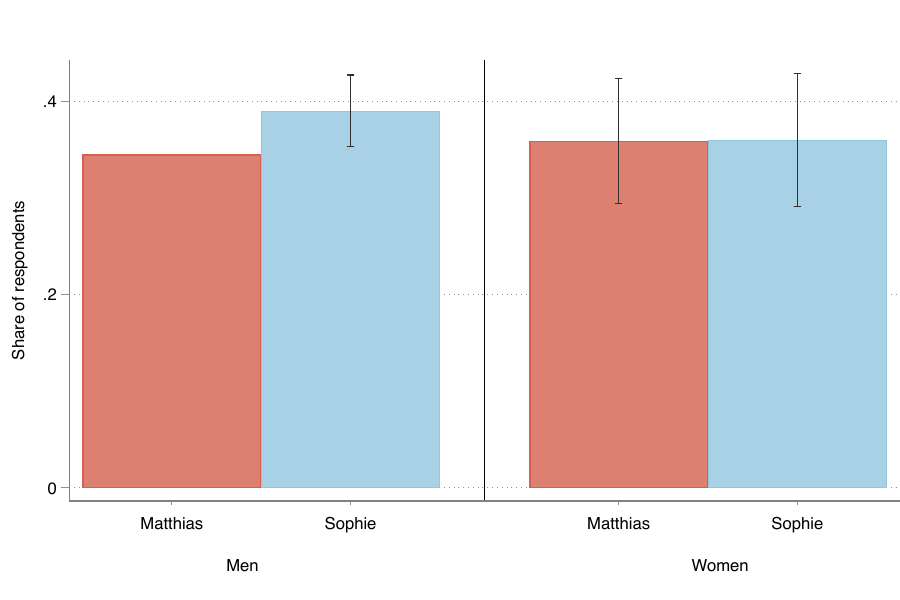}
	\end{minipage}	\\
		\vspace{1.2cm}
	\begin{minipage}[b]{1.0\linewidth}
		\centering
			\caption*{Panel B. Leader Won't Have a Good Experience}
		\includegraphics[scale=0.8]{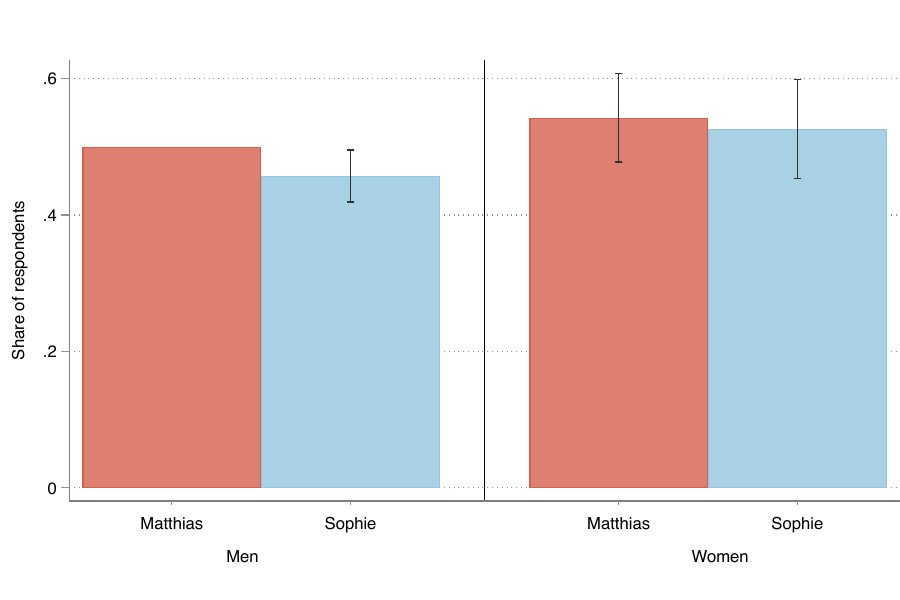}
	\end{minipage}	\\
		\vspace{0.5cm}
	\label{fig:vignette_leader}
	\begin{minipage}[b]{1.0\linewidth}
		\footnotesize \textit{Notes}: This figure illustrates leaders' perceptions about differential treatment  using a vignette approach based on the employee survey. The vignette introduced a fictional colleague whose first name was randomized to be either \textit{Matthias} or \textit{Sophie} to reflect a male or a female gender, respectively. Employees in leadership positions were asked to indicate how likely an employee would be to get hired if they were to apply for a team leadership position (Panel A) and whether they think they would have a good experience as a leader, especially with respect to  encountering conflict with team members or other leaders if they took on their first team leadership position (Panel B). Panel A shows the share of employees who report a high hiring probability, defined as 60\% or more. Panel B shows the share of employees who report expecting leaders won't have a good experience.  See Appendix Section \ref{sec:app_survey} for the exact question wording. Controls: Age, tenure, family status, hours, function.  N<3,xxx
		\end{minipage}	
\end{figure}

Men and women who currently hold leadership positions are also similar in how they describe the reality of being a leader. For instance, I find no gender differences in their  likelihood to report enjoying being a team leader or their self-reported likelihood of being an effective leader (Panel B of Appendix Figure \ref{fig:survey_leaders}). Similar patterns emerge from leaders' responses when presented with a vignette question that elicits career advice for a colleague with either a female name or a male name. Appendix Figure \ref{fig:vignette_leader} shows that I find no evidence that female leaders have more negative perceptions about their success likelihood or  treatment as leader. 

Taken together, these findings suggest that men and women who sort into leadership positions are similar in their (observable) career planning decisions and outcomes. This result is at stark contrast to the robust gender differences I find along these dimensions for employees in lower-level positions, suggesting that critical gender differences may occur early on in employees' careers. This contrasting finding is particularly important given that the vast majority of diversity policies target employees who already hold leadership positions, while very few policies target employees in lower-level positions (\citetalias{hkp2023}).

\begin{table}[h]
	\caption{Success Likelihood Conditional on Applying for a Promotion}
	\begin{center}
		\scalebox{1.0}{{
\def\sym#1{\ifmmode^{#1}\else\(^{#1}\)\fi}
\begin{tabular}{l*{2}{c}}
\hline\hline
                    &\multicolumn{1}{c}{Interview}&\multicolumn{1}{c}{Hiring}\\

 &\multicolumn{1}{c}{(1)}&\multicolumn{1}{c}{(2)}\\
\hline
Female              &       1.501&       1.425 \\
&     (0.541)         &     (0.772)         \\
\hline
Outcome Mean        &      0.4815         &      0.1650         \\
Av ME for Women     &      0.2722         &      0.1758         \\
Gender Gap in \%    &        56.5         &       106.5         \\
Observations          &        2xx               &        2xx               \\
\hline\hline \multicolumn{3}{p{0.6\linewidth}}{\footnotesize \textit{Notes}:  This table documents gender differences in success likelihood for employees in first leadership positions (hierarchy ranking between 20 and 40) who applied for an internal promotion. Promotions are defined using my preferred approach based on a combined measure of three dimensions of job authority (see Section \ref{sec:hierarchy}). Each coefficient stems from a separate logit regression of an indicator for getting invited for an interview or hired on gender and a large set of controls. Gender gaps in \% are computed by dividing the average marginal effect for women based on the logit coefficient by the outcome mean.  Controls: Age, German citizenship, educational qualification, marital status, family status, parental leave, firm tenure, function, division location, full-time, hours, number of direct reports, and quarter fixed effects. Standard errors in parentheses.}\\ \end{tabular}}
}
	\end{center}
	\label{table:hiring_top}
\end{table}

\FloatBarrier
\begin{figure}[!ht]
	\thisfloatpagestyle{plainlower}
	\caption{Gender Differences Among Leaders}
	\centering
		\vspace{0.2cm}
	\begin{minipage}[b]{0.8\linewidth}
		\centering
			\caption*{Panel A. Where would you like to see yourself with respect to your career?}
		\includegraphics[scale=2.4]{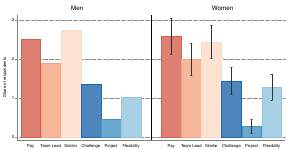}
	\end{minipage}	\\
		\vspace{1.2cm}
	\begin{minipage}[b]{1.0\linewidth}
		\centering
			\caption*{Panel B. Experience as Leader}
		\includegraphics[scale=2.5]{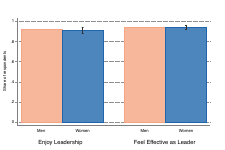}
	\end{minipage}	\\
		\vspace{0.5cm}
	\label{fig:survey_leaders}
	\begin{minipage}[b]{1.0\linewidth}
		\footnotesize \textit{Notes}: This figure shows gender differences for employees in leadership positions based on the employee survey. 
 Panel A shows answers to the question where employees would like to see themselves with respect to their career in five years. Employees could choose among following answers:  more pay, more responsibility over a team, staying in a similar job, more challenges, more project responsibility, and more job flexibility.  Panel B shows answers to questions about employees' self-perception as leaders. Bars 1 and 2 show the share of respondents who report enjoying leading a team. Bars 3 and 4 of Panel B show the share of respondents who report feeling that they are effective leaders.   Controls: Age, tenure, family status, hours, function.  N<3,xxx
		\end{minipage}	
\end{figure}
\FloatBarrier

\FloatBarrier
\FloatBarrier

\subsection{Constant Availability as Salient Feature of Team Leadership}
\label{sec:appendix_availability}
This section provides additional results on gender differences with respect to constant availability, which is also reported to be a salient feature of team leadership. In addition to the large administrative burden associated with leading large teams (named by 55\% of respondents) and the requirement to resolve conflict (named by 28\% of respondents), which are the two most commonly named features of team leadership that are unappealing to employees, 5\% of respondents name the constant availability that is required from team leaders as unappealing. For instance, one respondent describes the requirement as follows ``The different unexpected events that continuously happen on a frequent basis.''. Another respondent states ``The required constant availability on the part of the team and being responsible for everything require both professional and personal concessions.''. While the required availability likely scales with team size and therefore should be captured by my analysis of team size in Section \ref{sec:consequences}, this section provides separate results that focus on the availability requirement given its important role for gender differences documented in previous literature (e.g., \citealp{Goldin2014}, \citealp{MasPallais2017}, \citealp{wassermanhours}).

Using hypothetical job choice questions, I find that women are more likely than men to forgo higher pay to avoid positions that require unpredictable overtime. When asked to choose between a position that pays 30\% more than their current position and often requires overtime and a position that pays 10\% more and has very predictable hours, women are 25\% ($p$=0.000) less likely to choose the higher-paying position that requires overtime (Panel A of Appendix Figure \ref{fig:survey_constant}). This finding suggests that positions that require more unpredictable overtime, as team leadership positions likely do, may be less appealing to women. This difference could occur due to differences in preferences, constraints, or uncertainty about the costs that an employee may occur due to overtime.  In unreported results, I find that this gender gap is similar for employees who work full-time (and for whom the number of hours is likely less binding) and for employees who do not have children, suggesting that this gap is not simply explained by differences in part-time or family status. 

Men and women also differ in their perception of the extent to which team leader positions require constant availability. When asked how often they think team leaders experience unpredictable overtime, women in lower-level are 28\% ($p$=0.000) more likely to think that overtime occurs most days of the week (Bars 1 and 2 of Appendix Figure \ref{fig:survey_constant}, Panel B). This finding suggests that women relative to men in lower-level positions have more negative perceptions of the availability that team leadership requires, which may affect their decisions to apply.

However, in addition to this gender gap in perceptions among lower-level employees, I also find large gender differences in reported overtime among current leaders  (Bars 3 and 4 of Appendix Figure \ref{fig:survey_constant}, Panel B). Female leaders are 33\% ($p$=0.000) more likely to report overtime occurring most days a week. This finding implies that while women in lower-level positions have more negative perceptions than their male counterparts,  women's (more negative) perceptions relative to female leaders' reports are slightly more accurate than men's perceptions relative to male leaders' reports. In addition, both men and women underestimate the overtime that leaders face. 

How should we interpret the gender gap in reported overtime among leaders? One explanation that is consistent with this finding is that the reality of being a team leader may be different for female leaders. For instance, if female leaders differ in how they lead a team (e.g., extent of delegation, extent of mentoring,  non-promotable tasks), this type of leadership may require different hours. Similarly, if female leaders are approached differently by others (e.g., subordinates asking for help, supervisors delegating tasks), this may also yield longer hours. Another explanation is the fact that overtime and constant availability may be more salient to female leaders, making them more aware of this feature of leadership.

While a thorough test of these explanations is beyond the scope of this paper, suggestive evidence points rather towards the role of salience than to differences in the reality as leaders. In unreported results, I focus on the subsets of leaders who are less likely to face challenges with respect to their effectiveness: leaders with long tenures at the firm, those with a lot of experience as a leader, those with a high self-reported effectiveness as a leader,  and those with high confidence. Even in these subsamples, I find large gender differences in reported overtime, suggesting that (perceived) lower effectiveness as a leader is likely not the key driver. While women irrespective of their qualities may  feel the need to try harder if they perceive differential treatment, Panel B of Appendix Figure \ref{fig:vignette_leader} shows that in this setting, female leaders do not perceive female leaders to have a harder time relative to males. Instead, I find that large differences in reported overtime emerge for employees who have children, which is in line with overtime being more salient for mothers. 

Taken together, these findings suggest that to the extent to which overtime requirements could be reduced, team leadership positions could be made more appealing to women. Similarly, providing employees with advice on how to best respond to frequent overtime appears to be an effective policy implication. To the extent to which female leaders experience overtime differently from male leaders, such advice could be given in a gender-specific fashion. However, my results also suggest that in contrast to the large administrative burden associated with leading large teams and the requirement to resolve conflict, overtime is less important for explaining the application gap. While the other two facts jointly reduce the observed gender application gap for team leadership positions from 33.9\% to 18.2\%, adding controls for differences with respect to overtime reduces the application gap by less than 2 percentage points (Appendix Table \ref{table:applied_pref}). 

\FloatBarrier
\begin{figure}[!ht]
	\thisfloatpagestyle{plainlower}
	\caption{Gender Differences With Respect to Constant Availability}
		\centering
			\begin{minipage}[b]{0.8\linewidth}
			\centering
			\caption*{Panel A. Hypothetical Job Choice}
			\vspace{-4mm}
			\centering
			\includegraphics[scale=2.0]{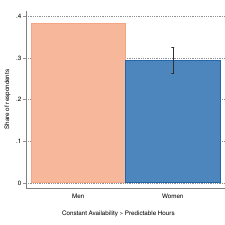}
		\end{minipage} \\	
		\vspace{6mm}
						\begin{minipage}[b]{0.8\linewidth}
		\centering
		\caption*{Panel B. Perception of Frequent Overtime}
		\vspace{-4mm}
		\centering
		\includegraphics[scale=2.0]{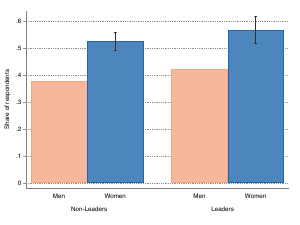}
	\end{minipage}
	\label{fig:survey_constant}
	\begin{minipage}[b]{0.8\linewidth}
		\footnotesize \textit{Notes}:  This figure illustrates gender differences with respect to the constant availability of team leaders based on the employee survey. Panel A illustrates gender differences in responses of employees in lower-level positions to a hypothetical choice question (N>5,xxx). The bars indicate the share of respondents who chose  a job that pays 30\% more and often requires overtime over a job that pays 10\% more and and has very predictable
hours. Panel B illustrates gender differences in employee perceptions of overtime. Bars 1 and 2 show the share of employees in lower-level positions who report that team leaders encounter unpredictable events
that lead them to work overtime most days of the week (N>5,xxx). Bars 3 and 4 show the corresponding share of current team leaders who report overtime occurring most days of the week (N<3,xxx). See Appendix Section \ref{sec:app_survey} for the exact question wording. Controls: Age, tenure, family status, hours, function.  
	\end{minipage}	
\end{figure}
\FloatBarrier

\subsection{External Validity: Survey Results from US Workers}
\label{sec:appendix_us}

To probe the robustness of my findings, I conducted an auxiliary survey with the white-collar and management employees of the firm that are based in the United States. The US survey used the same survey infrastructure as in Germany and received a 36.2\% response rate. 

The patterns that emerge in the US show striking similarities to the gender differences that I document in Germany. In the US, women in lower-level positions are 39\% ($p$=0.010) less likely to report wanting to lead a team in the next five years (Column 1 of Appendix Table \ref{table:gap_us}). When asked which type of position they would be interested in applying to in the next 12 months, women are 42\% ($p$=0.000) less likely to choose a higher-level position that requires leadership over a team (Column 2 of Appendix Table \ref{table:gap_us}).  Instead, they are more interested in higher-level positions that entail responsibility over a project or product and in expert roles (Panel B of Appendix Figure 	\ref{fig:career_aspiration_US}.)  Women are also 39\% ($p$=0.057) less likely to  have applied for a team leadership position (Column 3 of Appendix Table \ref{table:gap_us}).  

Similar to the results in Germany, I find no meaningful differences with respect to team leadership among employees who already hold leadership positions. Women and men are similarly likely to report wanting to take on more responsibility over a team (Columns 1 and 2 of Appendix Figure 	\ref{fig:career_aspiration_US_top}). I also find no meaningful gender differences in the extent to which team leaders enjoy leading a team (Columns 3 and 4 of Appendix Figure 	\ref{fig:career_aspiration_US_top}) and feel effective as team leaders (Columns 5 and 6 of Appendix Figure 	\ref{fig:career_aspiration_US_top}). 

The documented patterns resemble findings from surveys of US workers at other firms that document gender differences in wanting to manage others (\citetalias{HBR2014}). Taken together, these results suggest that the documented gender differences with respect to team leadership represent a broader phenomenon that is not restricted to the German context. 

One additional advantage of the US survey is that this setting allows me to test for the importance of other dimensions of diversity besides gender. To do so,  I elicit employees' self-reported ethnicity and race in the survey and define ethnic/racial minorities as employees who do not self-identify as White or those who do identify as Hispanic.\footnote{The exact question wording is: \textit{How would you describe your race/ethnicity? Please select all that apply. \{Hispanic/Latino, American Indian or Alaskan Native, Asian, Black or African American, Native Hawaiian or Other Pacific Islander, White / Caucasian, Other\}}. While my preferred specification pools all minority groups for power purposes, my results are robust to alternative comparison groups.} In contrast to the differences by gender, I do not find that minorities are less likely to want to lead a team (Appendix Table \ref{table:gap_us}), despite the fact that white employees represent the majority of the workforce in my sample, especially at higher levels. These results suggest that gender may play a special role when it comes to differences in team leadership.

\FloatBarrier
\begin{figure}[!ht]
	\thisfloatpagestyle{plainlower}
	\caption{Gender Differences Among US Employees in Lower-Level Positions}
	\centering
		\vspace{0.2cm}
	\begin{minipage}[b]{0.8\linewidth}
		\centering
			\caption*{Panel A. Where would you like to see yourself with respect to your career?}
		\includegraphics[scale=2.4]{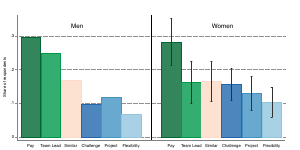}
	\end{minipage}	\\
		\vspace{1.2cm}
	\begin{minipage}[b]{1.0\linewidth}
		\centering
			\caption*{Panel B. Which position would you be most interested in applying to in the next 12 months?}
		\includegraphics[scale=2.5]{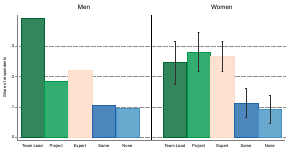}
	\end{minipage}	\\
		\vspace{0.5cm}
	\label{fig:career_aspiration_US}
	\begin{minipage}[b]{1.0\linewidth}
		\footnotesize \textit{Notes}: This figure shows gender differences with respect to career planning based on survey responses of the firm's lower-level employees that are based in the United States. Panel A shows answers to the question where employees would like to see themselves with respect to their career in five years. Employees could choose among following answers:  more pay, more responsibility over a team, staying in a similar job, more challenges, more project responsibility, and more job flexibility. Panel B shows answers to the question which positions employees would be most interested in applying to in the next 12 months. Employees could choose among following answers: Position at a higher level that requires leading a team, position at a higher level that requires responsibility over a product or project without responsibility over a team, position at a higher level  as an expert without responsibility over a team, position that is at a similar level as the current position, no interest in applying to any of these positions. 
 Controls: Age, tenure, family status, hours, function. N$<$ 1,XXX  
		\end{minipage}	
\end{figure}
\FloatBarrier

\FloatBarrier
\begin{figure}[!ht]
	\thisfloatpagestyle{plainlower}
	\caption{Gender Differences Among US Employees in Leadership Positions}
	\centering
		\vspace{0.2cm}
	\begin{minipage}[b]{0.8\linewidth}
		\centering
		\includegraphics[scale=2.4]{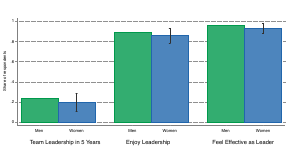}
	\end{minipage}	
	\label{fig:career_aspiration_US_top}
	\begin{minipage}[b]{1.0\linewidth}
		\footnotesize \textit{Notes}: This figure shows gender differences among employees in leadership positions that are based in the United States. Bars 1 and 2 show the share of employees who would like to see themselves in a position with more responsibility over a team in five years. Bars 3 and 4 show the share of respondents who report enjoying leading a team. Bars 5 and 6 show the share of respondents who report feeling that they are effective leaders.   Controls: Age, tenure, family status, hours, function. N$<$ 7XX
		\end{minipage}	
\end{figure}
\FloatBarrier

\begin{table}[h]
	\caption{Heterogeneity in Gender Differences With Respect to Responsibility Over a Team}
	\begin{center}
		\scalebox{0.9}{{
\def\sym#1{\ifmmode^{#1}\else\(^{#1}\)\fi}
\begin{tabular}{l*{3}{c}}
\hline\hline

                        &\multicolumn{1}{c}{Career Plan}&\multicolumn{1}{c}{Interest Applying}&\multicolumn{1}{c}{Applied}\\
   &\multicolumn{1}{c}{(1) }&\multicolumn{1}{c}{(2) }&\multicolumn{1}{c}{(3)}\\
                        \hline
        
                        Female                       &      -0.575&      -0.739&      -0.542   \\
                   &    (0.222)         &     (0.192)         &     (0.284)                     \\
                    Ethnic/racial minority    &       0.196         &       0.189         &       0.463  \\
                    &     (0.177)         &     (0.155)         &     (0.226)        \\
                    \hline
                    Outcome Mean       &      0.2287         &      0.3512         &      0.1232           \\
                    Av ME  for Women    &     -0.0888         &     -0.1478         &     -0.0479         \\
                    Av ME  for Minority & 0.0337         &      0.0400         &      0.0524           \\
                    Gender Gap in \%   &       -38.8         &       -42.1         &       -38.9         \\
                    Minority Gap in \% &        14.7         &        11.4         &        42.5       \\
                    Observations        &         9xx         &         9xx           &         9xx            \\

                    \hline\hline \multicolumn{4}{p{0.8\linewidth}}{\footnotesize \textit{Notes}:  This table documents differences in career planning and applications for lower-level employees based in the United States.  Column 1 focuses on wanting to lead a team in the next five years as outcome of interest. Column 2 focuses on being interested in applying for a higher-level position with team leadership as outcome of interest. Column 3  focuses on having applied for a team leadership position in the past 12 months as outcome of interest. Each coefficient stems from a separate logit regression of the given indicator on an indicator for gender, and indicator for ethnic/racial minority status, and a set of controls. Gaps in \% are computed by dividing the respective average marginal effect based on the logit coefficient by the outcome mean. Controls: Age, tenure, family status, hours, function. Standard errors in parentheses.}\\ \end{tabular}}
}
	\end{center}
	\label{table:gap_us}
\end{table}

\end{document}